\documentclass[fleqn,usenatbib]{mnras}

\usepackage{newtxtext,newtxmath}
\usepackage[T1]{fontenc}

\DeclareRobustCommand{\VAN}[3]{#2}
\let\VANthebibliography\thebibliography
\def\thebibliography{\DeclareRobustCommand{\VAN}[3]{##3}\VANthebibliography}

\usepackage{graphicx}\usepackage{subcaption}
\usepackage{amsmath}
\usepackage{CJKutf8}

\newcommand{\lps}{{\boldsymbol{\theta}_{\rm lp}}}

\title[Hierarchical Bayesian Inference of Globular Cluster]{Hierarchical Bayesian Inference of Globular Cluster Properties}

\author[R. Y. Wen et al.]{
Robin Y. Wen$^{1,2}$\thanks{E-mail: ywen@caltech.edu},
Joshua S. Speagle (\begin{CJK*}{UTF8}{gbsn}沈佳士\ignorespacesafterend\end{CJK*})$^{3,1,4,5}$,
Jeremy J. Webb$^{1}$, and
Gwendolyn M. Eadie$^{1,3}$
\\
$^{1}$David A. Dunlap Department of Astronomy \& Astrophysics, University of Toronto, 50 St George Street, Toronto ON M5S 3H4, Canada\\
$^{2}$California Institute of Technology, 1200 E. California Boulevard, Pasadena, CA 91125, USA\\
$^{3}$Department of Statistical Sciences, University of Toronto, 9th Floor, Ontario Power Building, 700 University Ave, Toronto, ON M5G 1Z5, Canada\\
$^{4}$Dunlap Institute for Astronomy \& Astrophysics, University of Toronto, 50 St George Street, Toronto, ON M5S 3H4, Canada\\
$^{5}$Data Sciences Institute, University of Toronto, 17th Floor, Ontario Power Building, 700 University Ave, Toronto, ON M5G 1Z5, Canada
}

\date{Accepted XXX. Received YYY; in original form ZZZ}

\pubyear{2023}

\begin{document}
\label{firstpage}
\pagerange{\pageref{firstpage}--\pageref{lastpage}}
\maketitle

\begin{abstract}
We present a hierarchical Bayesian inference approach to estimating the structural properties and the phase space center of a globular cluster (GC) given the spatial and kinematic information of its stars based on lowered isothermal cluster models. As a first step towards more realistic modelling of GCs, we built a differentiable, accurate emulator of the lowered isothermal distribution function using interpolation. The reliable gradient information provided by the emulator allows the use of Hamiltonian Monte Carlo methods to sample large Bayesian models with hundreds of parameters, thereby enabling inference on hierarchical models. We explore the use of hierarchical Bayesian modelling to address several issues encountered in observations of GC including an unknown GC center, incomplete data, and measurement errors. Our approach not only avoids the common technique of radial binning but also incorporates the aforementioned uncertainties in a robust and statistically consistent way. Through demonstrating the reliability of our hierarchical Bayesian model on simulations, our work lays out the foundation for more realistic and complex modelling of real GC data.
\end{abstract}


\section{Introduction}
Globular clusters (GCs) are massive collections of stars found in the outskirts of every type of galaxy. The dominant mechanisms that govern a cluster’s evolution --- stellar evolution, two-body relaxation, and the external tidal field of their host galaxy --- lead to clusters becoming mass segregated over time as they evolve towards a state of partial energy equipartition while playing host to stellar collisions and mergers \citep{87Spitzer,03HeggieStarCluster, Trenti2013}. These internal and external processes can also lead to clusters developing radial anisotropy \citep{Watkins2015, Tiongco2016, Jindal2019}. To gain a deeper understanding of how these various dynamical processes shape cluster evolution, we need to accurately measure the current spatial and kinematic distribution of stars within a given cluster, which can then be used to constrain the conditions under which GCs form. Ultimately, knowing the formation histories of a population of GCs provides insight into the formation and evolution of their host galaxy \citep{West2004}.

A large number of distribution functions (DFs) have been proposed to model the observed distribution of stellar positions and velocities in GCs \citep{54Woolley,63Michie,66King,75Wilson,15Gieles,19Claydon}. These DFs share the similarity that both density and velocity dispersion profiles decrease to zero out to a truncation radius. How the DF drops to zero varies from model to model, with additional treatments necessary to address radial anisotropy \citep{63Michie}, globular cluster rotation \citep{12VarriRotation}, or mass segregation \citep{76DaCostaMultiMass,15SollimaBias}.  

Historically, one can fit GCs with the aforementioned models by comparing the observed and theoretical surface brightness profiles or density profiles and sometimes kinematic profiles to determine the model parameters. Several different DF-based models have been fit to Galactic \citep{05McLaughlin,13Miocchi,19deBoerGCNumberDensityGaiaDR2,21Cohen_MW_InnerGC,23Cheng} and extragalactic GCs \citep{10WoodleyNGC5128,13WebbM87,13UsherNGC4278}. Beyond determining structural parameters, these DF-based models have also been used to probe more complex features of GCs such as the existence of intermediate-mass black holes (IMBHs) \citep{19Zocchi_IMBH} and their stellar initial mass functions (IMFs) \citep{23Dickson_IMF}. In addition to the traditionally-used surface  brightness, number density, and velocity dispersion profiles, the comparisons to DF-based models have increasingly considered other observables such as velocity anisotropy, stellar mass functions, and pulsar timings \citep{18Gieles_IMBH,20Brunet_47Tuc}.

The typical approach to fitting observational data with models is to first radially bin the observed stars, then estimate the uncertainties on the counts in each bin, and finally minimize the $\chi^2$ between the observed profile and the corresponding theoretical model profile. However, binning data and minimizing $\chi^2$ are undesirable for several reasons: 
\begin{enumerate}
    \item Information about each individual star is lost.
    \item Systematic errors are introduced due to the exact binning scheme used.
    \item Error estimation for the counts of each bin can be complex, depending on the completeness of the dataset, contamination from non-cluster stars, and measurement errors (which lead to inter-bin covariances).
    \item When trying to simultaneously fit multiple profiles, one must assume how to weight the importance of each fit. It must be decided whether the total $\chi^2$ is simply the sum of the individual $\chi^2$ values calculated for the density, kinematic and other profile fits or if they should be weighted differently, which is not a statistically robust procedure.
\end{enumerate}

To avoid radial binning and to make the inference more statistically robust, \citet{22EadieGCBayesianLIMEPY} (hereafter EWR22) has demonstrated the use of Bayesian inference to estimate the model parameters, cumulative mass profile, and mean-square velocity profile of a GC based on the positions and velocities of individual stars and assuming the lowered isothermal DF proposed in \citet{15Gieles} (hereafter GZ15). Focusing on the ideal cases with simulated data, EWR22 assumed that the positions and velocities of individual stars are completely known without any measurement errors, and their work investigated the catastrophic effects of selection bias on parameter fits if selection bias is not properly modelled.

The purpose of this work is to follow up on EWR22 to address the issues of incomplete data and measurement errors through hierarchical Bayesain inference (HBI). We build our hierarchical Bayesian model and demonstrate the use of HBI on simulated GC data to show the reliability and robustness of HBI to recover a cluster’s structural parameters in the presence of missing data and measurement uncertainties. Furthermore, we move away from the Cartesian coordinate used in EWR22 to the projected space on the plane of the sky (i.e., the reference frame in which actual data are measured), and we include the inference for the positions and velocities of the cluster center into our HBI model as well. The code to model GCs and perform inference with our hierarchical model can be found at \url{https://github.com/y52wen/hbmlimepy}.

The paper is organized as follows. In Sec.~\ref{sec:simulated-data}, we review the lowered isothermal DF in GZ15 and introduce the simulated dataset generated from their \texttt{limepy} python package that we use to validate our statistical method. We describe our hierarchical model in Sec.~\ref{sec:hierarchical-model} and specify our model priors in Sec.~\ref{sec:prior}, followed by a review of Hamiltonian Monte Carlo (HMC) in Sec.~\ref{sec:HMC} and a description of our approach to emulating the lowered isothermal DF through linear interpolation in Sec.~\ref{sec:emulation}, both of which are necessary for performing inference on our hierarchical model. In Sec.~\ref{sec:results}, we report our inference results on different GC simulations to demonstrate the reliability and robustness of our approach to model GCs. We summarize our findings and discuss future works for modelling real GC data in Sec.~\ref{sec:conclusions}.

\section{Simulated Data}\label{sec:simulated-data}
\subsection{The Lowered Isothermal DF}
In this work, we work with simulated data generated from the \texttt{limepy} code of lowered isothermal models for GCs introduced in GZ15. The \texttt{limepy} code has been successfully used to fit GCs from both real-world observation and simulated data. For example, \citet{16ZocchiN-body}, together with \citet{18GCNbody} and \citet{19Brunet_M4Nbody}, successfully demonstrated that direct N-body simulations of star clusters could be well-fit by the lowered isothermal DF, while \citet{19deBoerGCNumberDensityGaiaDR2} used this class of DF to determine the structure of galactic GCs using Gaia data \citep{18GaiaDR2-GC-DG}. The parameters for the lowered isothermal DF are $\boldsymbol{\theta}_{\rm lp}=(g,\Phi_0,M_{\rm tot},r_{\rm h})$, and the physical descriptions of these parameters are summarized in Table.~\ref{tab:sim-GC-param}. In general, $g$ and $\Phi_0$ impact the shape and concentration of the GC profile, while $M_{\rm tot}$ and $r_{\rm h}$ are scaling parameters that determine the mass and size of the GC. 

In this paper, we restrict our attention to isotropic GCs (i.e., the anisotropic radius $r_a\to\infty$ under the default in \texttt{limepy}) with only a single mass component (i.e. we assume stars in a GC have the same mass). In reality, GCs often develop some radial anisotropy due to various dynamical processes and GC stars have a wide range of masses, so it is better to treat a GC system as the combination of several single mass models \citep{76DaCostaMultiMass,19Brunet_M4Nbody}. However, anisotropic, multi-mass models pose significant challenges for hierarchical Bayesian inference, which we discuss in Sec.~\ref{sec:conclusions}, and so we leave its implementation to future work.

In the case of isotropic GCs, a value of $g = 0$ in the lowered isothermal model is equivalent to the \citet{54Woolley} model, while the $g = 1$ and the $g=2$ cases reduce to the \citet{66King} model and the \cite{75Wilson} non-rotating model respectively. Therefore, the lowered isothermal DF can be considered as the generalization of all of the above familiar DFs for GCs through varying $g$, the truncation parameter that decides how fast the DF falls off to zero at large radii. 

For the single-mass, isotropic models used in this work, the DF is given in GZ15:
\begin{align}
&f\left(E(r,v)\right)=\begin{cases}A E_\gamma\left(g,-\frac{E-\phi\left(r_{\mathrm{t}}\right)}{s^2}\right), & {\rm for}\; E \leqslant \phi\left(r_{\mathrm{t}}\right)\\
0\;, & {\rm for}\; E>\phi\left(r_{\mathrm{t}}\right)
\end{cases}\label{eq:DF}\\
&E(r,v)=v^2 / 2+\phi(r)\label{eq:E}
\end{align}
The DF in Eq.~\ref{eq:DF} depends on the total energy $E$ in Eq.~\ref{eq:E}, where $v$ is the velocity and $\phi(r)$ is the potential at distance $r$ from the centre. In Eq.~\ref{eq:DF}, the energy $E$ is lowered by the potential at the truncation radius $\phi\left(r_{\mathrm{t}}\right)$. The function $E_\gamma(g, x)$ is defined as
\begin{align}
E_\gamma(g, x)= \begin{cases}\exp (x), & g=0 \\ \exp (x) P(g, x), & g>0\label{eq:Egamma}\end{cases},
\end{align}
where $P(g, x) \equiv \gamma(g, x) / \Gamma(g)$ is the regularized lower incomplete gamma function (see Appendix~D of GZ15 for properties of $E_\gamma(g, x)$ and $P(g, x)$). The potential function $\phi(r)$ in Eq.~\ref{eq:DF} and \ref{eq:E} needs to be solved consistently through the following non-linear Poisson's equation 
\begin{equation}
\frac{1}{r^2}\frac{d}{dr}\left(r^2\frac{d\phi}{dr}\right)=4\pi G\rho,\;{\rm where}\; \rho=\int d^3v f(E(r,v)) \label{eq:Poisson},
\end{equation}
where DF $f(E(r,v))$ given in Eq.~\ref{eq:DF} also depends non-linearly on $\phi(r)$. To solve Eq.~\ref{eq:Poisson} numerically, the equation is often non-dimensionalized as in \citet{66King} and GZ15. 

In the dimensionless case, the model is completely specified by two parameters: the central potential $\Phi_0$, which is a required boundary condition for solving Poisson's equation and defines how concentrated the model is; and the truncation parameter $g$, which controls the sharpness of the truncation of the model. The physical units of the model are defined by two scales: the velocity scale $s$ and the phase-space normalization constant $A$, which in turn determine the total mass $M_{\rm total}$ and the GC radius scale $r_{\rm h}$. These two scale parameters $A$ and $s$ in Eq.~\ref{eq:DF} need to be solved consistently with the Poisson's Equation of Eq.~\ref{eq:Poisson} to ensure proper normalization of the DF to 1, so $A$ and $s$ are in fact functions of $\boldsymbol{\theta}_{\rm lp}$. The \texttt{limepy} code fully implements and solves the above Eq.~\ref{eq:DF}-\ref{eq:Poisson} and is able to give the DF $f(r,v)$ for any $\boldsymbol{\theta}_{\rm lp}$.

\subsection{Simulation Parameters}
\begin{table}
	\centering
	\begin{tabular}{|l|l|l|} 
		\hline
		$\mathbf{\theta}_{\rm lp}$ & Description &  Possible Values\\
		\hline
		$g$ & truncation parameter  & $1.2,1.6,2.0$\\
		$\Phi_0$ & central gravitational potential  & $3.0,5.0,8.0$\\
		$M_{\rm tot}$ & total mass ($M_{\odot}$) & $10^5,10^6$\\
        $r_{\rm h}$ & half-light radius (pc) & $3.0,9.0$\\
		\hline
	\end{tabular}
 	\caption{\texttt{Limepy} model parameter values used to simulate stars in GCs. We consider all possible combinations of these four globular cluster parameters. The parameter combination $(g=2.0,\Phi_0=8.0)$ is excluded due to its closeness to the cutoff boundary (shown in Fig.~\ref{fig:rvrh_cutoff}) that distinguishes realistic GC models from unrealistic ones (see Sec.~\ref{sec:prior} for a more detailed explanation). This leaves us with 32 sets of different parameterss. We generate 10 simulations for each of the parameter sets and test our model on these simulations.}
	\label{tab:sim-GC-param}
\end{table}

\begin{table}
	\centering
	\begin{tabular}{|l|l|l|} 
		\hline
		$\mathbf{\theta}$ & Description &  Possible Values\\
		\hline
        $N$ & number of stars measured in cluster  & $100,500,1000$\\
        $R_{\rm c}$ & distance of cluster center to earth (kpc) & $1,2,5,10$ \\
        $\sigma$ & measurement uncertainties (mas, mas/year) & $0.02, 0.1, 0.5$\\
		\hline
	\end{tabular}
 	\caption{Hyperparameter values used to simulate stars in GCs that reflect realistic observation conditions. We fix the structural GC parameters at $(g,\Phi_0,M_{\rm tot},r_{\rm h})=(2.0,5.0,10^5,3)$ while changing the above hyper-parameters of the simulations individually. In the third row, $\sigma$ denotes the measurement errors for angular positions (mas), parallax (mas), and proper motions (mas/year) for an individual star, where we assume the errors for all five components are the same. We also assume that every star in GC shares the same measurement uncertainties.}
	\label{tab:sim-obs-param}
\end{table}

We develop and test our method for GC parameter inference with simulated spatial and kinematic data of GC stars in the Heliocentric equatorial coordinate system. The \texttt{limepy} code can directly sample stars from the lowered isothermal DF in Cartesian coordinates $(\mathbf{x}_{\rm GC},\mathbf{v}_{\rm GC})$ within a GC-centric reference frame, and we translate the stars' coordinates to the Heliocentric Cartesian coordinate $(\mathbf{x},\mathbf{v})$ with the GC center parameters $\boldsymbol{\theta}_c$ (the default value given in Eq.~\ref{eq:default-center}). We then convert the Heliocentric Cartesian coordinates $(\mathbf{x},\mathbf{v})$ to the Heliocentric equatorial coordinates $(\boldsymbol{q},\boldsymbol{p})$, for which the transformations $(\boldsymbol{q},\boldsymbol{p})=T^{-1}(\mathbf{x},\mathbf{v})$ are explicitly given in Eq.~\ref{eq:R}-\ref{eq:vR} of Appendix~\ref{sec:coordinate}.

To demonstrate the reliability and the robustness of our method, we test and validate our HBI model on different structural parameters of GCs covering a wide range of GC morphology. For the values given in Table~\ref{tab:sim-GC-param}; we consider all possible combinations of these four globular cluster parameters (excluding parameters with $(g=2.0,\Phi_0=8.0)$ due to the cutoff boundary shown in Fig.~\ref{fig:rvrh_cutoff}), which result in 32 sets of parameters with different values. We generate 10 simulations, with each simulation being a different realization of the lowered isothermal DF, for every parameter set. We will then obtain constraints though our HBI model on these 320 simulations. By default, we draw $N=1000$ stars from the lowered isothermal DF in each simulation. In the Gaia DR2 data, the number of stars with available proper motion in a Galactic GC can range from tens to thousands \citep{19Vasiliev_GaiaDR2_pm}, with 1000 being a representative figure that balances the amount of information available and the complexity of the model to demonstrate the performance of our inference procedure.

The default position parameters for the centers of globular clusters are chosen to be
\begin{align}
    \boldsymbol{\theta}_{c}&=(\boldsymbol{q}_{\rm c},\boldsymbol{p}_{\rm c})=(\alpha_{\rm c},\delta_{\rm c},\pi_{\rm c},\mu_{\alpha*,{\rm c}},\mu_{\delta,{\rm c}},v_{R,{\rm c}})\nonumber\\
    &=(\frac{\pi}{3} {\, \rm rad},\frac{\pi}{4} {\,\rm rad}, 1{\,\rm mas},4 {\,\rm mas/year}, 5 {\,\rm mas/year}, 30 {\,\rm km/s})\label{eq:default-center}
\end{align}
in the heliocentric equatorial coordinate where $\mu_{\alpha*}=\mu_{\alpha}\cos\delta$. Here the angular centers are arbitrarily chosen, and the angular velocities of the clusters are chosen to reflect Cartesian velocities in tens of km/s \citep{21VasilievGaiaDR3GC}. The choice of $R_{\rm c}$ impacts the accuracy of the constraints, since GCs located further away have more uncertain distance estimates. We assume the cluster is observed at a distance of $R_{\rm c} =1$ kpc, which corresponds to the parallax value $\pi_c=1\;{\rm mas}$. In reality, GCs can have distances of a few kpcs or tens of kpcs away, and we investigate the impacts of changing $R_{\rm c}$ values in Sec.~\ref{sec:hyper-param}.

We also add measurement errors to the simulated data $\boldsymbol{d}=(\boldsymbol{q},\boldsymbol{p})$ for each star. The position and proper motion measurements $\boldsymbol{d}^{\rm obs}$ for each star are assumed to follow a normal distribution with a certain standard deviation $\sigma$ (which we refer to as the measurement error). Based on \cite{18Gaia_summary}, we assume the measurement errors for the declination angle $\alpha$, the right ascension $\delta$, and the parallax $\pi$ for all stars are $\sigma_{\alpha}=\sigma_{\delta}=\sigma_{\pi}=0.1$ mas, and the measurement errors for the proper motions are assumed to be $\sigma_{\mu_{\alpha*}}=\sigma_{\mu_{\delta}}=0.1$ mas/year. The above assumed astrometric measurement errors correspond to assuming our samples stars with G-magnitude at around 17 \citep{18Gaia_summary}. By default, we assume that the radial velocities for all stars are unknown, which is a reasonable assumption for cluster stars measured through the Gaia data.

In each case, we assume that there are 1000 stars with measured properties in each cluster. However, we also explore the effects of different numbers of star as given in the first row of Table~\ref{tab:sim-obs-param} (see also Sec.~\ref{sec:hyper-param}) on the inference results. We also explore the impact of different measurement uncertainties (as given in the third row of Table~\ref{tab:sim-obs-param}) in the angular positions and angular velocities on the parameter inference. When we run the 320 simulations with different lowered isothermal model structural parameters $(\Phi_0,g,M_{\rm tot},r_{\rm h})$, we keep all the hyperparameters in Table~\ref{tab:sim-obs-param} at their default values $(R_{\rm c},N_{v_{\rm R}},\delta, N)=(1\;{\rm kpc},0,0.1,1000)$. When we explore the impacts of different hyperparameter values, we keep the structural parameters fixed at $(\Phi_0,g,M_{\rm tot},r_{\rm h})=(5,2,10^5,3)$.

\section{Methods}\label{sec:methods}

\subsection{A Hierarchical Model}\label{sec:hierarchical-model}

\begin{figure}
	\includegraphics[width=\columnwidth]{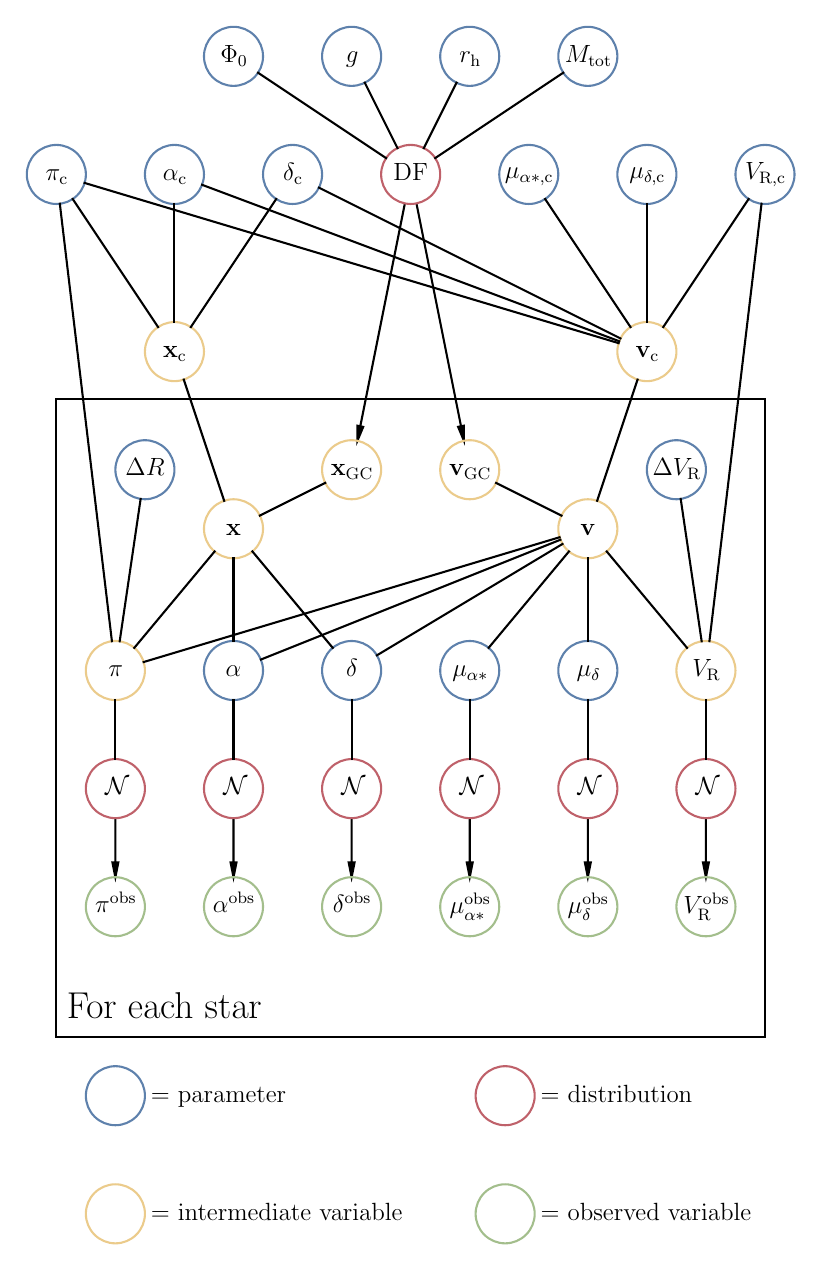}
    \caption{Graphical representation of the hierarchical Bayesian model built in this work to analyze simulated GC data with the lowered isothermal distribution function. All six phase-space measurements in the heliocentric sky coordinates are given a multilevel treatment; the observed variables are assumed to be drawn from Gaussian distributions centered around true (latent) phase-space parameters, with standard deviations equal to the measurement errors. For the positions and proper motions, the measurement uncertainties from Gaia are incorporated into the model. The heliocentric latent phase-space parameters are then transformed into the Cartesian coordinate and used to estimate the four structural parameters of the lowered isothermal distribution function and the six phase-space parameters that describe the GC center. For the parallax and the radial velocity of the individual star, we sample in $\Delta R=R-R_{\rm c}$ and $\Delta V_{\rm R}=V_{\rm R}-V_{\rm R,c}$) --- that is, the difference between the star's radius (radial velocity) and those of the GC center. We use these differences as the sampling parameter to reduce the correlations between parameters and improve the performance of HMC.}
    \label{fig:HBI-graph}
\end{figure}

Using the simulated spatial and kinematic data of stars from each GC mentioned in Sec.~\ref{sec:simulated-data}, we take a hierarchical Bayesian approach to infer the model parameters. With the maturation of inference techniques for large, complex Bayesian models over the last two decades, hierarchical modelling has gained momentum and been applied to a wide range of scientific problems across many disciplines. In the context of astronomy, HBI has been used to estimate photometric redshift in galaxy surveys \citep{16LeistedtHBI-redshift,23LeistedtHBI-redshift-SPS}, and estimate the mass of the Milky Way in \citep{18Eadie_MUGS2,19EadieGCMW,22ShenMWMassHB}, to name a few examples. A graphical representation for our hierarchical model to determine GC properties is shown in Fig.~\ref{fig:HBI-graph}. 

From Bayes' theorem, the posterior probability of a vector of model parameters $\boldsymbol{\theta}$ given data $\boldsymbol{d}$, is
\begin{equation}
p(\boldsymbol{\theta}|\boldsymbol{d})=\frac{p(\boldsymbol{d}|\boldsymbol{\theta})p(\boldsymbol{\theta})}{p(\boldsymbol{d})}\propto p(\boldsymbol{d}|\boldsymbol{\theta})p(\boldsymbol{\theta})\label{eq:Bayesian}
\end{equation}
where $p(\boldsymbol{d}|\boldsymbol{\theta})$ is the probability of the data conditional on the model parameters (also known as the likelihood $\mathcal{L}(\boldsymbol{\theta}; \boldsymbol{d})$, a function of model parameters for fixed data), $p(\boldsymbol{\theta})$ is the prior distribution on the model parameters, and $p(\boldsymbol{d})$ is the "evidence" or prior predictive density. The latter is a constant, leaving us with a target distribution proportional to the posterior distribution $p(\boldsymbol{\theta}|\boldsymbol{d})$, which we estimate through Markov Chain Monte Carlo (MCMC) sampling (see Sec.\ref{sec:HMC}). We devote the rest of this section to specifying our model, which is summarized with the directed acyclic graph (DAG) shown in Fig.~\ref{fig:HBI-graph}.

Our simulated data products, described in Sec.~\ref{sec:simulated-data}, are the six observed Heliocentric equatorial phase-space components of each individual star 
\begin{equation*}
\boldsymbol{d}^{\rm obs}_i=(\boldsymbol{q}^{\rm obs}_i,\boldsymbol{p}^{\rm obs}_i)=(\alpha^{\rm obs}_i,\delta^{\rm obs}_i,\pi^{\rm obs}_i,\mu_{\alpha*,i}^{\rm obs},\mu_{\delta,i}^{\rm obs}v_{\rm R,i}^{\rm obs}),
\end{equation*}
which are shown in the base level of the graphical representation for our Bayesian model in Fig.~\ref{fig:HBI-graph}.

Each component of a star's observed equatorial phase-space coordinates $d^{\rm obs}$ \footnote{For notation simplicity, we will hereafter drop the index $i$ and simply use the bold font $\boldsymbol{d}\equiv\boldsymbol{d}_i$ to represent the phase-space coordinates of one star whenever such practice does not cause confusion. The non-bold symbol $d$ will be used to denote a single component of the phase-space coordinates of a GC star.} is assumed to follow the normal distribution 
\begin{equation}
    p(d^{\rm obs}|d)\sim\mathcal{N}(d,\sigma_{d})\label{eq:measurement-error-modelling},
\end{equation}
where $d$ is the individual component of the true heliocentric equatorial phase-space coordinates $(\boldsymbol{q},\boldsymbol{p})$), and $\sigma_{d}$ is the observed measurement error for the particular phase-space component of the star. We have assumed the measurements for each phase-space component are independent, which is often not the case for Gaia measurements since the measurements for parallax, angular position, and proper motions are often correlated with non-negligible covariance. Here we assumed independent Gaussian distributions for simplicity, and we can potentially change to multivariate Gaussian distributions to incorporate covariance. Any missing data can be incorporated as a variable with a flat (uniform) prior. Our model therefore maximizes the available phase-space information. We are able to make parameter estimates using any combination of sources, whether they are missing positions, velocities, neither, or even both.

The true heliocentric equatorial phase-space components $(\boldsymbol{q},\boldsymbol{p})$ of each star are then transformed into the Heliocentric Cartesian coordinates $(\mathbf{x},\mathbf{v})=T(\boldsymbol{q},\boldsymbol{p})$, where the coordinate transformation $T$ is given in Eq.~\ref{eq:x}-\ref{eq:vz} of Appendix~\ref{sec:coordinate}. We next transform the Heliocentric coordinates into the GC-centric coordinates for each star:
\begin{equation}
(\mathbf{x}_{\rm GC},\mathbf{v}_{\rm GC})=(\mathbf{x},\mathbf{v})-(\mathbf{x}_{\rm c},\mathbf{v}_{\rm c})=T(\boldsymbol{q},\boldsymbol{p})-T(\boldsymbol{q}_{\rm c},\boldsymbol{p}_{\rm c}),\label{eq:xv-GC}
\end{equation}
where the center of GC in Heliocentric Cartesian coordinates is $(\mathbf{x}_{\rm c},\mathbf{v}_{\rm c})=T(\boldsymbol{q}_{\rm c},\boldsymbol{p}_{\rm c})=T(\boldsymbol{\theta}_{\rm c})$. For Eq.~\ref{eq:xv-GC}, we assume there is no rotation present for stars around GC centers. In the future, we could include rotation for more complex GC modelling (see \citet{19SollimaGCRotation} for an example). 

As seen in Fig.~\ref{fig:HBI-graph}, we treat the the coordinates of the GC center $\boldsymbol{\theta}_{\rm c}$ as parameters and directly infer them from our hierarchical model. We therefore obtain estimates for GC position and velocity centers in a Bayesian framework, which is in contrast to the usual approach of setting the mean of GC stars' positions as the GC center \citep{18GaiaDR2-GC-DG,19deBoerGCNumberDensityGaiaDR2} and other more sophisticated approaches to determine centers such as the pre-slice method and the density contour method \citep{10Goldsbury_ACS}. The angular positions of GCs can be treated as fixed; each star's angular position has been measured with exquisite accuracy through Gaia. However, for the distance and kinematic centers of GCs, our Bayesian method can robustly include measurement errors and missing data for stellar position and kinematic data, with the potential for providing better and more statistically consistent constraints on GC phase-space centers. Our inference results on the GC centers for our simulations are discussed in detail in Sec.~\ref{sec:results}.

Given the GC structural parameters $\boldsymbol{\theta}_{\rm lp}$, the likelihood for observed GC stars having true positions and velocities $(\vec{\mathbf{x}}_{\rm GC},\vec{\mathbf{v}}_{\rm GC})$\footnote{Here the bold vector notation $\vec{\mathbf{x}}_{\rm GC}$ is used to represent the Cartesian position coordinates of all stars in contrast to the coordinates of one star $\mathbf{x}_{\rm GC}\equiv \mathbf{x}_{\rm GC,i}$.} (in GC-centric Cartesian coordinates) is
\begin{equation}
p(\vec{\mathbf{x}}_{\rm GC},\vec{\mathbf{v}}_{\rm GC}|\boldsymbol{\theta}_{\rm lp})=\prod_{i=1}^N \frac{f_{\rm lp}\left(\mathbf{x}_{\rm GC,i},\mathbf{v}_{\rm GC,i}|\boldsymbol{\theta}_{\rm lp}\right)}{M_{\text {total }}},
\label{eq:limepy-df-likelihood}
\end{equation}
where the lowered isothermal DF $f_{\rm lp}$ is specified in Eq.~\ref{eq:DF}, $N$ is the number of stars, and the DF is normalized by the total mass to become a probability distribution function. By taking the products of DF evaluated at each star, Eq.~\ref{eq:limepy-df-likelihood} assumes that the sampling of observed stars from all GC stars are uniform and each observed star is independently selected, which is an overtly idealistic assumption for real surveys. We will try to incorporate more realistic modelling for selection functions in future works.

To write the DF in Eq.~\ref{eq:limepy-df-likelihood} in terms of the true Heliocentric equatorial coordinates, we use the coordinate transformations in Eq.~\ref{eq:xv-GC} and obtain
\begin{align}
p(\vec{\mathbf{q}},\vec{\mathbf{p}}|\boldsymbol{\theta}_{\rm lp},\boldsymbol{\theta}_{\rm c})&=\prod_{i=1}^N \frac{f_{\rm lp}\left(T(\boldsymbol{q}_i,\boldsymbol{p}_i)-T(\boldsymbol{q}_{\rm c},\boldsymbol{p}_{\rm c})|\boldsymbol{\theta}_{\rm lp}\right)}{M_{\text {total }}}\nonumber\\
&\quad\cdot\prod_{i=1}^N\left\lvert\frac{\partial T(\boldsymbol{q}_i,\boldsymbol{p}_i)}{\partial(\boldsymbol{q}_i,\boldsymbol{p}_i)}\right\rvert,
\label{eq:limepy-df-likelihood-phase-space}
\end{align}
where the second term is the Jacobian for the coordinate transformations from the Heliocentric equatorial coordinates $(\vec{\mathbf{p}},\vec{\mathbf{q}})$ to the GC-centric Cartesian coordinates $(\vec{\mathbf{x}}_{\rm GC},\vec{\mathbf{v}}_{\rm GC})$. The translation by the GC center does not change the Jacobian. Ignoring the Jacobian terms in Eq.~\ref{eq:limepy-df-likelihood-phase-space} will bias the inference results, especially for the distance of the cluster center $R_{\rm c}$.

For the Heliocentric equatorial coordinates of each star, the parallax (distance) is known with the least accuracy among the position coordinates, while the radial velocity can be unknown. As a result, $\pi_{i}$ and $\pi_{\rm c}$ (and $V_{R,i}$ and $V_{R,c}$) are highly correlated with each other, since a shift in the cluster center $\pi_{\rm c}$ (or $V_{R,c}$) can be compensated by a reverse shift in $\pi_{i}$ (or $V_{R,i}$) for all stars. To reduce the correlations among parameters for better inference performance, we define the new parameters $\Delta R_i=R_i-R_{\rm c}$ and $ \Delta V_{R,i}=V_{R,i}-V_{R,{\rm c}}$. In these new definitions, $\Delta R_{i}$ and $\Delta V_{R,i}$ are centered at zero when no radial distances or radial velocities of the stars are known, and these variables for individual stars will no longer shift along with the values of the cluster centers. We will infer $\Delta R_i$ and $\Delta V_{R,i}$ instead of $R_i$ and $V_{R,i}$. For convenience, we write the coordinate parameters in which we perform inference as 
\begin{equation}
    (\boldsymbol{s},\boldsymbol{t})=(\alpha,\delta,\Delta R,\mu_{\alpha*},\mu_{\delta},\Delta V_{\rm R})\label{eq:sampling-parameter-definition},
\end{equation}
which are circled in blue in Fig.~\ref{fig:HBI-graph}. The relationship between the sampling coordinates and the Heliocentric equatorial coordinates is
\begin{align}
    \left(\boldsymbol{q},\boldsymbol{p}\right)&=\left(\boldsymbol{q}(\boldsymbol{s}),\boldsymbol{p}(\boldsymbol{t})\right)\nonumber\\
    &=\left(\alpha,\delta,A/(\Delta R+A/\pi_{\rm c}),\mu_{\alpha*},\mu_{\delta},\Delta V_{\rm R}+V_{\rm R,c}\right),\label{eq:sampling-parameter}
\end{align}
where $A=1\,{\rm kpc}/1\,{\rm mas}$ is the conversion factor between distance and parallax. 

For the error modelling of the individual star given in Eq.~\ref{eq:measurement-error-modelling},
\begin{equation}
p\left(\vec{\boldsymbol{q}}^{\rm obs},\vec{\boldsymbol{p}}^{\rm obs}|\vec{\boldsymbol{q}}(\vec{\boldsymbol{s}}),\vec{\boldsymbol{p}}(\vec{\boldsymbol{t}})\right)=p\left(\vec{\boldsymbol{q}}^{\rm obs},\vec{\boldsymbol{p}}^{\rm obs}|\boldsymbol{\theta}_{\rm lp},\boldsymbol{\theta}_{\rm c},\vec{\boldsymbol{s}},\vec{\boldsymbol{t}}\right)\label{eq:individual-level}
\end{equation}
since $(\vec{\boldsymbol{q}}^{\rm obs},\vec{\boldsymbol{p}}^{\rm obs})$ only depends on the GC parameters $(\boldsymbol{\theta}_{\rm lp},\boldsymbol{\theta}_{\rm c})$ through the sampling coordinates $(\vec{\boldsymbol{s}},\vec{\boldsymbol{t}})$. The lowered isothermal DF written in terms of the Heliocentric equatorial coordinates is given in Eq.~\ref{eq:limepy-df-likelihood-phase-space}, which models the relationship between each individual star and the entire GC. Using Bayes' theorem, we can now write down the posterior of our whole model, which is illustrated in Fig.~\ref{fig:HBI-graph}:
\begin{align}
    &p\left(\boldsymbol{\theta}_{\rm lp},\boldsymbol{\theta}_{\rm c},\vec{\boldsymbol{s}},\vec{\boldsymbol{t}}|\vec{\boldsymbol{q}}^{\rm obs},\vec{\boldsymbol{p}}^{\rm obs}\right)\nonumber\\
    &\propto p\left(\vec{\boldsymbol{q}}^{\rm obs},\vec{\boldsymbol{p}}^{\rm obs}|\boldsymbol{\theta}_{\rm lp},\boldsymbol{\theta}_{\rm c},\vec{\boldsymbol{s}},\vec{\boldsymbol{t}}\right) p\left(\boldsymbol{\theta}_{\rm lp},\boldsymbol{\theta}_{\rm c},\vec{\boldsymbol{s}},\vec{\boldsymbol{t}}\right)\nonumber\\
&=p\left(\vec{\boldsymbol{q}}^{\rm obs},\vec{\boldsymbol{p}}^{\rm obs}|\vec{\boldsymbol{q}}(\vec{\boldsymbol{s}}),\vec{\boldsymbol{p}}(\vec{\boldsymbol{t}})\right)p\left(\vec{\boldsymbol{s}},\vec{\boldsymbol{t}}|\boldsymbol{\theta}_{\rm lp},\boldsymbol{\theta}_{\rm c}\right)p\left(\boldsymbol{\theta}_{\rm lp},\boldsymbol{\theta}_{\rm c}\right)\label{eq:model-likelihood-simple}\\
&=\prod_{i=1}^N \Bigg\{p\left(\vec{\boldsymbol{q}}^{\rm obs}_i,\vec{\boldsymbol{p}}^{\rm obs}_i|\vec{\boldsymbol{q}}_i(\vec{\boldsymbol{s}}_i),\vec{\boldsymbol{p}}_i(\vec{\boldsymbol{t}}_i)\right) \nonumber\\
&\qquad \frac{f_{\rm lp}\left(T\left(\boldsymbol{q}_i(\boldsymbol{s}_i),\boldsymbol{p}_i(\boldsymbol{t}_i)\right)-T\left(\boldsymbol{q}_{\rm c},\boldsymbol{p}_{\rm c}\right)|\boldsymbol{\theta}_{\rm lp}\right)}{M_{\text {total }}} \left\lvert\frac{\partial T\left(\boldsymbol{q}_i(\boldsymbol{s}_i),\boldsymbol{p}_i(\boldsymbol{t}_i)\right)}{\partial(\boldsymbol{s}_i,\boldsymbol{t}_i)}\right\rvert\Bigg\}\nonumber\\
&\qquad p\left(\boldsymbol{\theta}_{\rm lp},\boldsymbol{\theta}_{\rm c}\right)\label{eq:model-likelihood},
\end{align}
where Eq.~\ref{eq:model-likelihood-simple} is obtained using Eq.~\ref{eq:individual-level} and the definition of the conditional probability, and Eq.~\ref{eq:model-likelihood} is obtained by substitution with Eq.~\ref{eq:limepy-df-likelihood-phase-space}. The Jacobian term in the second line of Eq.~\ref{eq:model-likelihood} is explicitly specified by Eq.~\ref{eq:Jacobian-s-t} of Appendix~\ref{sec:coordinate}. Eq.~\ref{eq:model-likelihood} reflects the hierarchical structure of our model: the first line of Eq.~\ref{eq:model-likelihood} gives the individual-level (star-level) modelling, while the second line specifies the population-level (GC-level) modelling, and they together constitute our model likelihood. The third line $p\left(\boldsymbol{\theta}_{\rm lp},\boldsymbol{\theta}_{\rm c}\right)$ is our model prior, which we will discuss in the following section.

\subsection{Prior Distributions}\label{sec:prior}

\begin{figure}
\begin{center}
	\includegraphics[width=0.8\columnwidth]{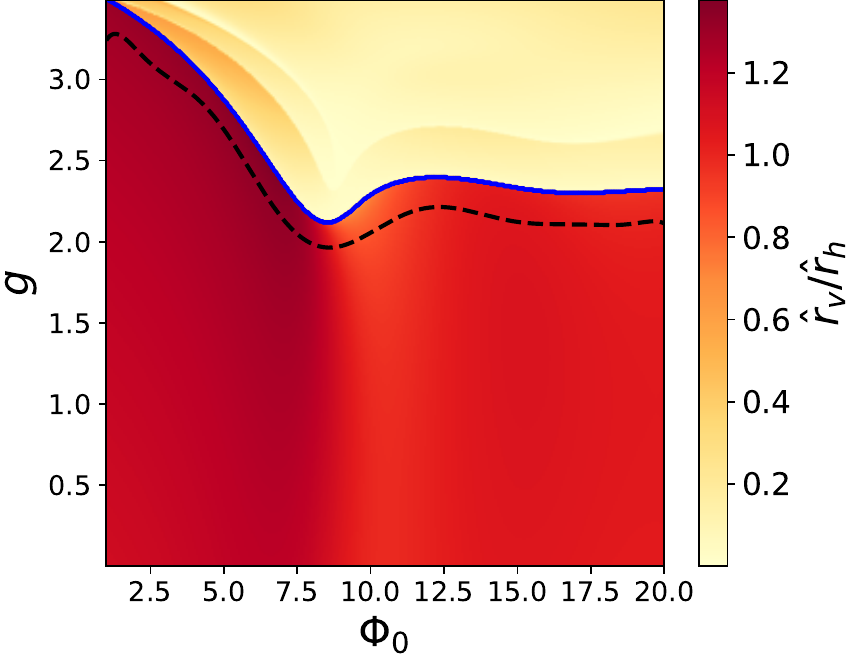}
\end{center}
    \caption{Ratio of the dimensionless virial radius to the half-mass radius, $\frac{\hat{r}_v}{\hat{r}_h}$ for models with different $\Phi_0$ and $g$. The blue line shows the boundary corresponding to $\hat{r}_v/\hat{r}_h\gtrapprox  0.64$, which distinguishes the lowered isothermal models (in the red region) that are relevant to modeling GCs. The black dashed-line represents the $10^{\rm th}$-degree polynomial $F_b(\Phi_0)$ that we used as the upper boundary in the uniform prior of $g$ for our Bayesian model. Notice that our prior function $F_b(\Phi_0)$ cuts off some physically relevant ($\Phi_0$, $g$) parameters close to the blue boundary.}
    \label{fig:rvrh_cutoff}
\end{figure}

Two advantages of Bayesian inference are the ability to incorporate meaningful prior information and the necessity to state this explicitly. Our goal is to choose priors that are broad and generally non-restrictive so that we can apply the same priors to models with a wide parameter range, while excluding parameters regions that are physically impossible or irrelevant. We now clarify our choice of the model prior $p\left(\boldsymbol{\theta}_{\rm lp},\boldsymbol{\theta}_{\rm c}\right)$, which has the following structure:

\begin{align} 
p(\boldsymbol{\theta}_{\rm lp},\boldsymbol{\theta}_{\rm c})&=p(\boldsymbol{\theta}_{\rm lp})p(\boldsymbol{\theta}_{\rm c})\\
p(\boldsymbol{\theta}_{\rm lp})&=p(\Phi_0,g)p(r_{\rm h})p(M_{\rm tot})\\
p(\boldsymbol{\theta}_{\rm c})&=p(\alpha_{\rm c})p(\delta_{\rm c})p(\pi_{\rm c})p(\mu_{\alpha*,{\rm c}})p(\mu_{\delta,{\rm c}})p(V_{\rm R,{\rm c}})\label{eq:prior-center},
\end{align}
where the priors for all the GC parameters are independent except for $\Phi_0$ and $g$. For the structural parameters $\boldsymbol{\theta}_{\rm lp}$, we have some prior knowledge on the total mass $M_{\rm total}$ and half-mass radius $r_{\rm h}$ based on observations of Milky Way GCs, while having considerably less prior information on the values of $g$ and $\Phi_0$, aside from the physically allowable, positive values. We choose the priors for $g$ and $\Phi_0$ as the following:
\begin{align}
\Phi_0 &\sim \operatorname{Unif}(1.5,14)\label{eq:prior-Phi0}, \\
g &\sim \operatorname{Unif}(0.001,F_b(\Phi_0))\label{eq:prior-g}.
\end{align}
Our prior for $\Phi_0$ is the same as the one in EWR22, while we modify the prior on the truncation parameter $g$ in EWR22 by imposing an upper boundary $F_b(\Phi_0)$ that depends on $\Phi_0$ to exclude models that are inapplicable to GCs. As discussed in \citet{14GomezTruncatedExponential} and GZ15, the density profile for the lowered isothermal model is infinite when $g\geq 3.5$. There is also a class of isotropic models that are finite in extent but showing a sharp upturn in the density profile at large radii, which makes these models inapplicable to star clusters. To separate these inapplicable models from those realistic models for GCs, we follow GZ15 and use \texttt{limepy} to compute the ratio of the dimensionless virial radius $\hat{r}_v =-G \hat{M}^2/(2 \hat{U})$ over $\hat{r}_h$ for a grid of models with $1\leq \Phi_0\leq 20$ and $0.001\leq g \leq 3.49$ (Fig.~\ref{fig:rvrh_cutoff}). One can see that for any given $\Phi_0$ value, once $g$ increases beyond a certain threshold, the change in $\hat{r}_v/\hat{r}_h$ becomes abrupt, which signals the separation between two classes of models. We use $\hat{r}_v/\hat{r}_h> 0.64$ as the criterion for identifying the class of models relevant to modelling GCs. The upper boundary $F_b(\Phi_0)$ function for $g$ is chosen to be a $10^{\rm th}$-degree polynomial fitted to the boundary corresponding to $\hat{r}_v/\hat{r}_h\approx 0.64$ (the blue line in Fig.~\ref{fig:rvrh_cutoff}), shifted downward by $0.2$. $F_b(\Phi_0)$  is plotted as the black dashed line in Fig.~\ref{fig:rvrh_cutoff}. The 0.2 downward shift is chosen to exclude parameter spaces that are too close to the $\hat{r}_v/\hat{r}_h\approx 0.64$ boundary. The black dashed-line represents the  $F_b(\Phi_0)$ that we used as the upper boundary in the uniform prior of $g$ for our Bayesian model. 

For the mass and radius of GCs, we sample the parameters and set the priors in log scale where $\log_{10}M_{\rm total}$ and $\log_{10}r_{\rm h}$ are assumed to follow normal distributions (in units of pc and $M_{\odot}$ respectively):
\begin{align}
\log_{10}M_{\rm total} &\sim N\left(\mu_M=5.85, \sigma_M=0.6\right)\label{eq:prior-M},\\
\log_{10}r_{\rm h} &\sim N\left(a=0,b=1.5,\mu_r=0.7,\sigma_r=0.3\right)\label{eq:prior-rh},
\end{align}
where $\log_{10}r_{\rm h}$ is truncated by the lower bound $a$ and the upper bound $b$ to ensure the GC radius is both positive and not overly large ($r_{\rm h}\lessapprox 30\,{\rm pc}$). Our prior for $M_{\rm total}$ is the same as the one chosen in EWR22. In general, GC masses span about an order of magnitude and  setting the log-normal prior on $M_{\rm total}$ is supported by the near universal GC mass function \citep{10HarrisGCCatalogue}. For $r_{\rm h}$, EWR22 chose a more restrictive prior that also depends on the value of $r_{\rm h}$, while we choose a broader prior that reflects the distributions for the half-mass radii of 168 Milky Way GCs included in the Baumgardt's catalogue \footnote{\url{https://people.smp.uq.edu.au/HolgerBaumgardt/globular/}}, of which the structural parameters are determined in \citet{18GCNbody}. In general, our inference results are not sensitive to the exact choice of priors for $M_{\rm total}$ and $r_{\rm_h}$.

For the GC center parameters $\boldsymbol{\theta}_{\rm c}$, we assume the prior for each phase-space component is independent as seen in Eq.~\ref{eq:prior-center}. For each component $d_{\rm c}$ (where $d=\alpha,\delta,\pi,\mu_{\alpha*},\mu_{\delta},V_{\rm R}$), we adopt an empirical Bayes prior and assume
\begin{equation}
    d_{\rm c}\sim \operatorname{Unif}\left({\rm min}\{d^{\rm obs}_i|i=1..N\},{\rm max}\{d^{\rm obs}_i|i=1..N\}\right)
    \label{eq:prior-center-specific}
\end{equation}
such that the component of the center is bounded by the minimum and maximum of the corresponding components of all observed stars in GCs. The above prior ensures the basic physical validity of the model while remaining generally uninformative. One can possibly put a tighter prior on the GC center compared to Eq.~\ref{eq:prior-center-specific}. 

For the coordinate parameters of each individual star $(\vec{\boldsymbol{s}},\vec{\boldsymbol{t}})$ (Eq.~\ref{eq:sampling-parameter-definition}), which we also need to sample in MCMC, they are entirely determined by the conditional probability $p\left(\vec{\boldsymbol{s}},\vec{\boldsymbol{t}}|\boldsymbol{\theta}_{\rm lp},\boldsymbol{\theta}_{\rm c}\right)$, which is considered as part of the likelihood in Eq.~\ref{eq:model-likelihood}. From a sampling perspective, $(\vec{\boldsymbol{s}},\vec{\boldsymbol{t}})$ can be considered as having a flat (non-informative) prior. For practical purposes, flat priors are hard to sample, so we recast the measurement error distribution $p(d^{\rm obs}|d)$ in Eq.~\ref{eq:measurement-error-modelling} (the individual-level modelling) as the prior for $\alpha,\delta,\mu_{\alpha*},\mu_{\delta}$ and remove these distributions from the likelihood part of Eq.~\ref{eq:model-likelihood}. For $\Delta R$ and $\Delta V_{\rm R}$, the priors are still chosen as flat and $p(\pi^{\rm obs}|\pi)$ and $p(V_{\rm R}^{\rm obs}|V_{\rm R})$ remain as part of the likelihood due to our change of sampling coordinates from $\pi$ and $V_{\rm R}$ to $\Delta R$ and $\Delta V_{\rm R}$. In general, the measurement error modelling distributions $p\left(\vec{\boldsymbol{q}}^{\rm obs},\vec{\boldsymbol{p}}^{\rm obs}|\vec{\boldsymbol{q}},\vec{\boldsymbol{p}}\right)$ in Eq.~\ref{eq:model-likelihood-simple} has the flexibility to be either treated as part of the likelihood or the prior for the true phase-space coordinates $(\vec{\boldsymbol{q}},\vec{\boldsymbol{p}})$ in MCMC. 

\subsection{Inference with HMC}\label{sec:HMC}

The standard approach of Bayesian inference is to use Markov Chain Monte Carlo (MCMC) algorithms, where the posterior $p(\boldsymbol{\theta}|\boldsymbol{d})$ in Eq.~\ref{eq:Bayesian} is directly sampled based on the likelihood and the priors without the need to perform integration to determine the evidence $p(\boldsymbol{d})$. At each step, the MCMC algorithm makes proposals to new positions in parameter space and checks whether the posterior at the proposed position is favorable relative to the current position. 

The likelihood and priors of our hierarchical model have been specified in Sec.~\ref{sec:hierarchical-model} and \ref{sec:prior}. In total, there are $6N+10$ parameters (Fig.~\ref{fig:HBI-graph} and Eq.~\ref{eq:model-likelihood}), where $N$ is the number of observed stars in the GC. Traditional MCMC algorithms such as Metropolis-Hastings and Gibbs samplers \citep{Metropolis,Hastings,92CasellaGibbs} are unable to perform inference for high-dimensional problems with hundreds or thousands of parameters, which is the case for our model. Other more modern MCMC samplers such as affine-invariant ensemble sampler and slice sampler also struggle in high-dimensional problems \citep{15HuijserAffineInvariantSampler,Slice}. For hierarchical models with analytical probability distributions in the exponential family, it is often possible to marginalize over certain lower-level parameters (or nuisance parameters) to significantly reduce the dimensions of problems \citep{15Eadie,BDAGelman}. However, for our model likelihood in Eq.~\ref{eq:model-likelihood}, the DF is based on a numerical ordinary differential equation (ODE) solver implemented in GZ15 without analytical solutions. Furthermore, the coordinate transformations from $(\vec{\boldsymbol{q}},\vec{\boldsymbol{p}})$ to $(\vec{\boldsymbol{x}}_{\rm GC},\vec{\boldsymbol{v}}_{\rm GC})$ are highly non-trivial, which prevents the marginalization over the phase space coordinates $(\vec{\boldsymbol{q}},\vec{\boldsymbol{p}})$ of each individual star. Therefore, our inference problem remains intrinsically high-dimensional and intractable with traditional MCMC methods.

Hamiltonian Monte Carlo (HMC) is a gradient-based MCMC algorithm that avoids random-walk behavior by making better proposals using Hamiltonian dynamics \citep{87DuaneHMC,11NealHMC}. By using the gradient of the log-posterior to inform the movement of the particle, HMC avoids staying in the same position for long periods due to bad proposals, and ensures that subsequent proposals are distant in parameter space. HMC provides numerous benefits over traditional MCMC algorithms, particularly in rending high-dimensional inference problems feasible due to its fast, scalable algorithm. For a more detailed introduction to HMC and a discussion of its benefits, we refer the reader to \citet{BDAGelman} and \citet{17BetancourtHMCIntro}. 

For the practical use of HMC into inference problems, there are two main challenges. First of all, HMC requires the gradient of the likelihood with respect to all inference parameters, which is computationally expensive and not always available in the analytical form, although automatic differentiation \citep{15BaydinAD} makes this tractable. We will discuss our approach to rendering our model likelihood automatically differentiable in Sec.~\ref{sec:emulation}. The second drawback of HMC is that the gradient, through the geometry of the target distribution depends on the specific parameterization of the problem. This dependence means that the step size, the Euclidean metric (which accounts for linear correlations in the posterior), and the number of steps in each HMC iteration need to be tuned manually to ensure good and efficient sampling of the parameter space. The tuning of these hyper-parameters is highly non-trivial and also model dependent. In the worst case of a poorly tuned HMC, the particle may repeatedly loop back in a way that subsequent samples are very close together, which is known as the “U-Turn problem" \citep{14HoffmanNUTS}.

The No-U-Turn sampler (NUTS) is an adaptive extension of HMC with several self-tuning strategies that remove the need to manually select HMC parameters  \citep{14HoffmanNUTS}. As a result, the end user normally only needs to worry about the design of their model and not about computational inefficiencies or the detailed implementation of the NUTS itself. It is the advent of NUTS and its efficient implementation in probabilistic programming packages like \texttt{PyMC} \citep{16SalvatierPyMC} and \texttt{Stan} \citep{17Stan, Rstan} that make the inference of our hierarchical model and other recent works \citep[e.g.,][]{22ShenMWMassHB} a tractable problem.

In this work, we make use of \texttt{PyMC}, which is an open-source probabilistic programming package written in Python for Bayesian inference. It implements many inference algorithms including the aforementioned Metropolis-Hasting sampling, slice sampling \cite{Slice}, and NUTS. Treating the NUTS implemented in \texttt{PyMC} as a black-box can be insufficient for certain models, since the self-tuning strategies adopted in NUTS are not always guaranteed to work. The specific parametrization of the problem and the details in the numerical implementation can still impact or even determine the quality and efficiency of the sampling process. As discussed in Sec.~\ref{sec:hierarchical-model}, we choose $\Delta R$ and $\Delta V_{\rm R}$ as the sampling parameters instead of $\pi$ and $V_{\rm R}$ to decorrelate the radial coordinates of individual stars from those of the GC center. If we sample in $\pi$ and $V_{\rm R}$, the sampled posteriors of these highly-correlated radial coordinates behave badly with additional nonphysical oscillatory features, though the mean and variance of the estimated posteriors remain sensible, which motivates our use of $\Delta R$ and $\Delta V_{\rm R}$. In addition, there are certain cases when the inference of our model fails with the NUTS implemented in \texttt{PyMC}, which we will discuss in Sec.~\ref{sec:result-GC-structure} and \ref{sec:conclusions}.

\subsection{Emulation of the lowered isothermal DF}\label{sec:emulation}

\begin{figure*}
\begin{subfigure}{.49\textwidth}
\centering
	\includegraphics[width=\columnwidth,height=4.26cm]{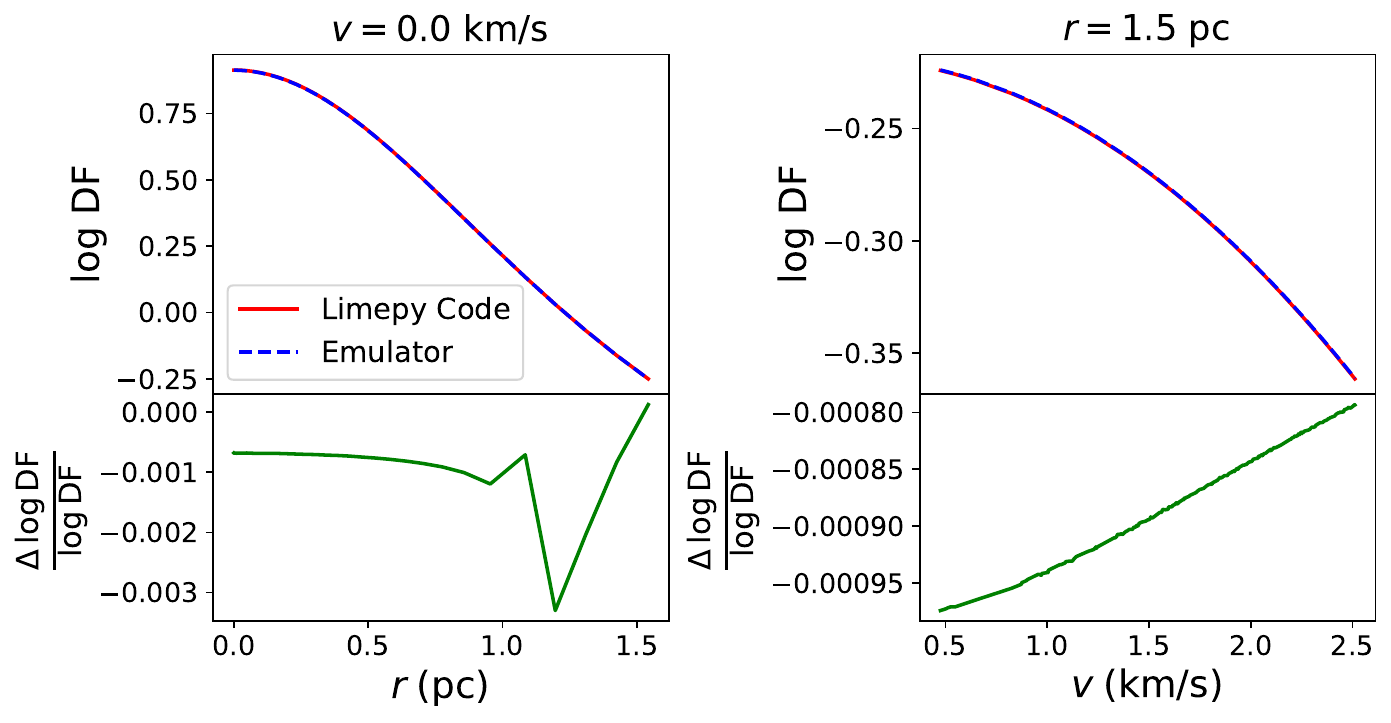}
     \caption{$(\Phi_0,g,\log_{10}M_{\rm tot},r_{\rm h})=(1.5,0.002,4,1)$}
     \label{fig:emulator-accuracy-1041}
 \end{subfigure}
\hfill
\begin{subfigure}{.49\textwidth}
\centering
 \includegraphics[width=\columnwidth,height=4.26cm]{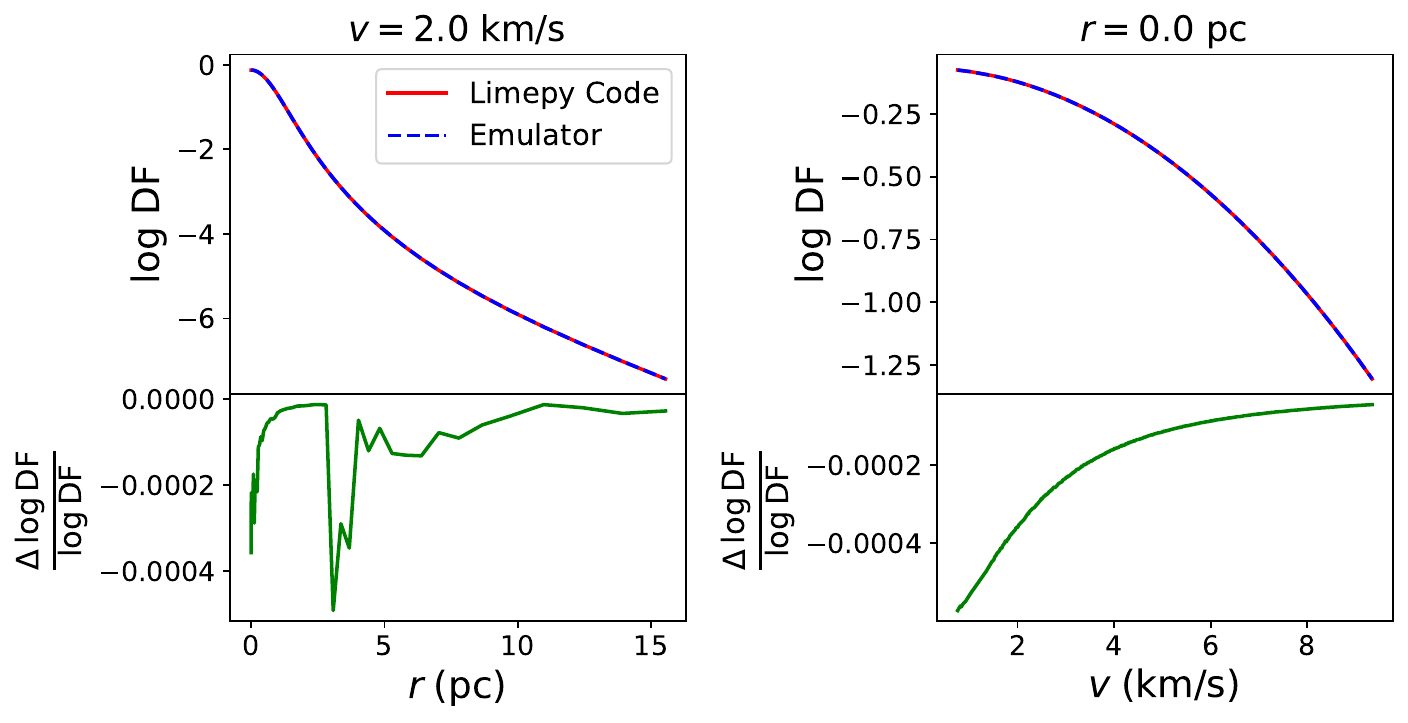}
   \caption{$(\Phi_0,g,\log_{10}M_{\rm tot},r_{\rm h})=(5,2,5,3)$}
   \label{fig:emulator-accuracy-5253}
\end{subfigure}

\begin{subfigure}{.49\textwidth}
\centering
	\includegraphics[width=\columnwidth,height=4.26cm]{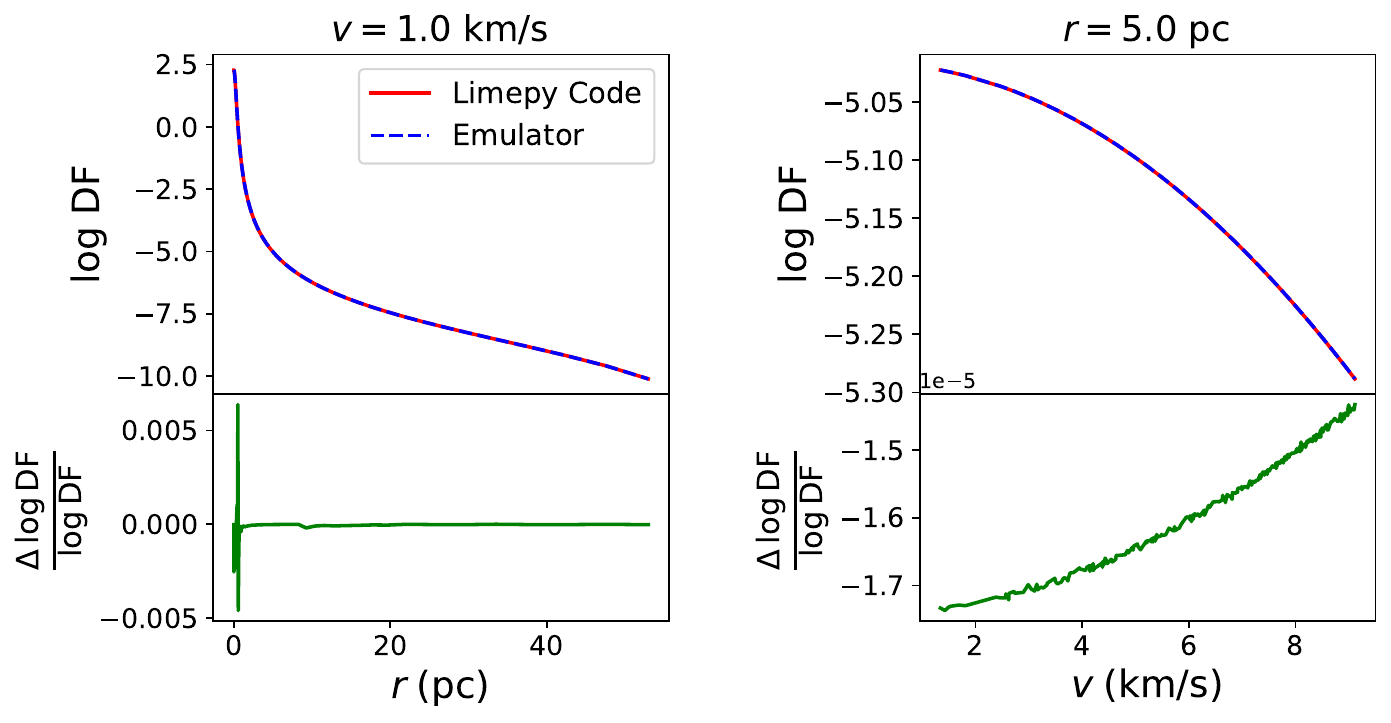}
     \label{fig:emulator-accuracy-10169}
      \caption{$(\Phi_0,g,\log_{10}M_{\rm tot},r_{\rm h})=(10,1,6,9)$}
\end{subfigure}
\hfill
\begin{subfigure}{.49\textwidth}
\centering
 \includegraphics[width=\columnwidth,height=4.26cm]{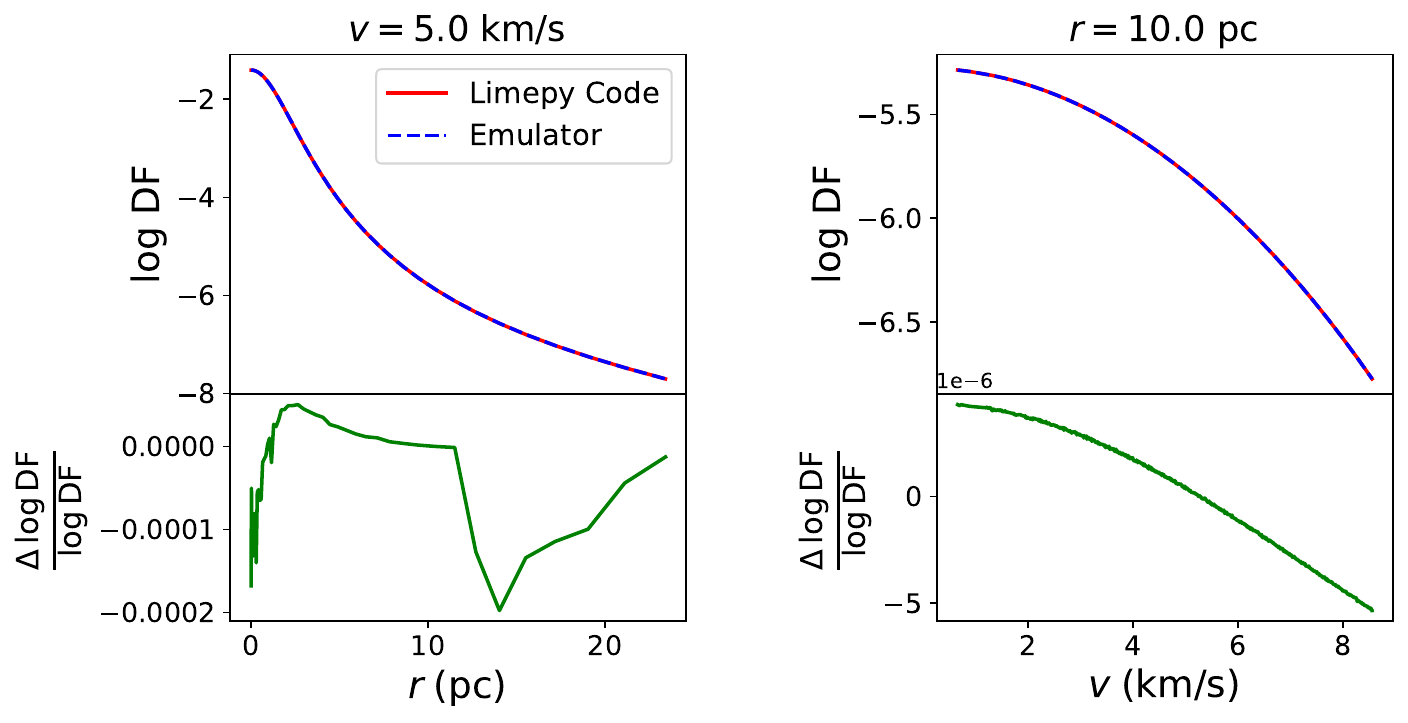}
  \label{fig:emulator-accuracy-70612}
   \caption{$(\Phi_0,g,\log_{10}M_{\rm tot},r_{\rm h})=(7.5,0.5,5.5,12)$}
\end{subfigure}
\caption{The evaluations of the lowered isothermal DF $f(r,v)$ in Eq.~\ref{eq:DF} under four different parameter combinations (subplots a, b, c, and d) by both the original \texttt{limepy} code (red solid line) and our interpolation-based emulator (blue dashed line) introduced in Sec.~\ref{sec:emulation}. The four parameter combinations are indicated by the titles of the subplots and are chosen to represent a broad range of values for $\boldsymbol{\theta}_{\rm lp}$. For visualization, we fix $v$ at a certain value in the left panels to show the variation of the DF with respect to $r$, and we fix $r$ in the right panels to show the change of DF with respect to $v$. We plot the values of DF in the log scale, which is the scale used for computation in MCMC, in the upper panels. The relative errors of $\log f$, which are defined as $\Delta \log f/\log f\equiv (\log f_{\rm emulator}-\log f_{\texttt{limepy}})/\log f_{\texttt{limepy}}$, are shown in the lower panels. In general, we see excellent agreement between the two methods in all four parameter combinations. Our emulator has an error smaller than $0.1\%$ in most of the parameter regions for $\log f$. The relative error only exceeds $0.1\%$  when the true $\log f$ values are close to 0.}
\label{fig:emulator-accuracy}
\end{figure*}

\begin{figure*}
\begin{subfigure}{.47\textwidth}
\centering
	\includegraphics[width=0.9\columnwidth]{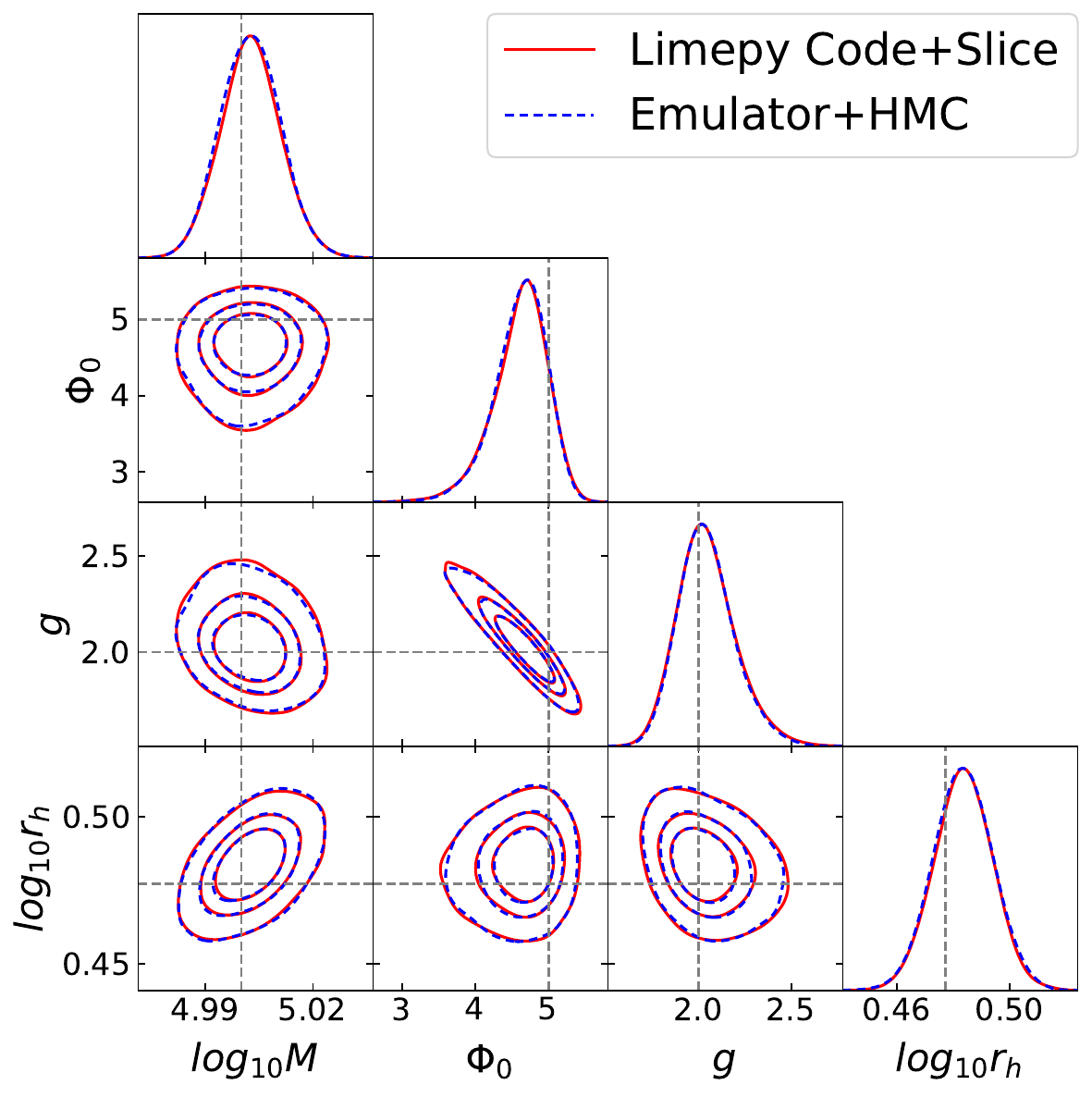}
     \label{fig:basic_Bayesian_200}
     \caption{True parameters $(\Phi_0,g,\log_{10}M_{\rm tot},r_{\rm h})=(5,2,5,3)$}
 \end{subfigure}
\hfill
\begin{subfigure}{.47\textwidth}
\centering
 \includegraphics[width=0.9\columnwidth]{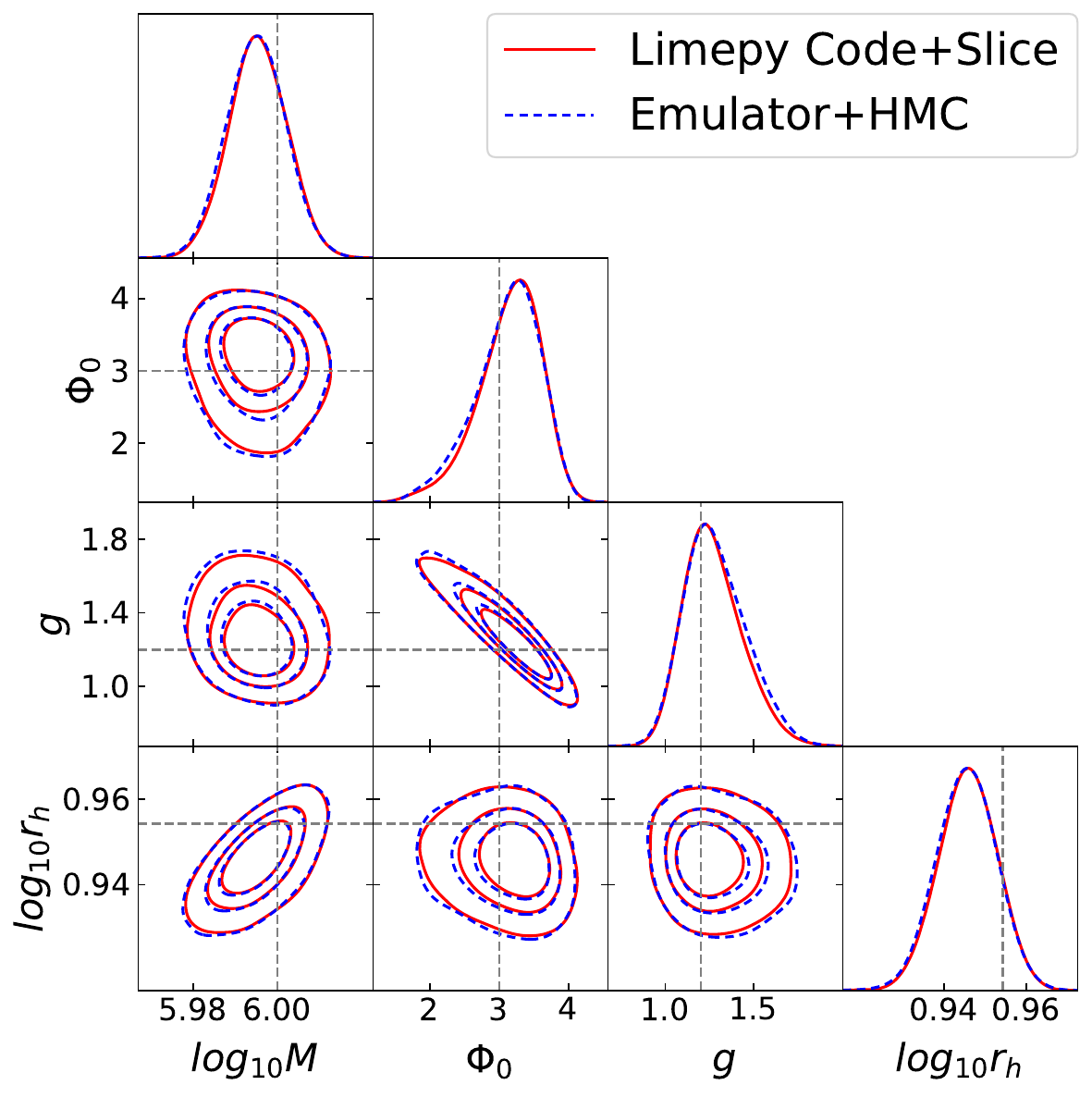}
  \label{fig:basic_Bayesian_70}
   \caption{True parameters $(\Phi_0,g,\log_{10}M_{\rm tot},r_{\rm h})=(3,1.2,6,9)$}
\end{subfigure}
\caption{Bayesian inferences of $(\Phi_0,g,\log_{10}M_{\rm tot},\log_{10}r_{\rm h})$ GC structural parameters for two different GC simulations of 1000 stars generated using the \texttt{limepy} code. For each simulation, we assume the 3D radius and velocity of each star from the GC center are completely and accurately known without any measurement errors. The red contours (showing $50\%$, $75\%$, and $95\%$ credible regions) are the inference results obtained with HMC sampling, and the likelihood is evaluated using our interpolation emulator of the lowered isothermal DF. The blue contours are obtained with Slice Sampling, and the likelihood is evaluated using the original \texttt{limepy} code. The dashed line gives the underlying correct parameter value used for generating the simulations. We see that the two runs with different DF evaluation methods and sampling methods agree with each other for both simulations, which suggests that our emulator is sufficiently accurate for performing Bayesian inference.} 
\label{fig:basic_Bayesian}
\end{figure*}

As discussed in the previous section, HMC requires the gradient of the likelihood with respect to all model parameters, which can be difficult and expensive to obtain for functions without simple analytical derivatives. For our model likelihood in Eq.~\ref{eq:model-likelihood}, it is difficult to obtain derivatives with respect to terms involving the lowered isothermal DF, which is obtained through the \texttt{limepy} code by numerically solving the Poisson Equation. For every different set of $\boldsymbol{\theta}_{\rm lp}$, we have to resolve the Poisson Equation, which is a time-consuming process for performing inference using MCMC. Obtaining reliable numerical derivatives on limepy DF through simple methods like finite difference is challenging for multi-parameter numerical functions. Luckily, automatic differentiation (AD) allows one to surpass this challenge and obtain reliable derivatives for complicated numerical programs. 

In general, AD techniques can transform a program that calculates numerical values of a function into a program that calculates numerical values for derivatives of that function, with about the same accuracy and efficiency as the function values themselves \citep{00BiggsAD,15BaydinAD}. Using AD requires one to write the numerical program in specialized packages and frameworks, and in this work, we use \texttt{JAX} \citep{18JAX}, a Python framework capable of automatically differentiating native Python and NumPy functions. In addition, \texttt{JAX} allows just-in-time (JIT) compilation \citep{XLA} of Python codes, which can significantly speed up the evaluation of Python programs. In our case, JIT allows quicker evaluation of our model likelihood in Eq.~\ref{eq:model-likelihood}, thereby speeding-up the HMC inference process. 

To render the \texttt{limepy} DF differentiable, we would therefore need to rewrite the \texttt{limepy} code in \texttt{JAX}, which is a highly non-trivial task requiring major code restructuring. The latter would be needed to eliminate branching, and significant efforts would be required to ensure the reliability of gradients under numerical ODE solvers. This is beyond the scope of this work. However, for the single-mass, isotropic limepy models, we can instead build a simple emulator of the lowered isothermal DF through linear interpolation, which we describe next.

In the DF for the single-mass, isotropic model (Eq.~\ref{eq:DF}), $A$, $r{\rm_t}$, $s$, and $\phi$ need to be determined by solving Poisson's equation (Eq.~\ref{eq:Poisson}). Note that $A$, $r{\rm_t}$, and $s$ are functions of $\boldsymbol{\theta}_{\rm lp}$, while $\phi$ is a function of both $\boldsymbol{\theta}_{\rm lp}$ and $r$. These functions only depend on the size $r_{\rm h}$ and mass $M_{\rm total}$ of GCs through the following scaling relations:
\begin{equation}
    r_{\rm t}\sim r_{\rm h},\;s^2\sim GM_{\rm total}/r_{\rm h},\; A\sim M_{\rm total}/(s^3 r_{\rm h}^3), \phi\sim s^2\label{eq:scaling}
\end{equation}
Therefore, to bypass the ODE solver, we
\begin{enumerate}
    \item linearly interpolate $\hat{A}(\Phi_0,g)$, $\hat{r}_{\rm t}(\Phi_0,g)$, $\hat{s}(\Phi_0,g)$, and $\hat{\phi}(\Phi_0,g,r)$ evaluated at $r_{\rm h,ref}=1\,{\rm pc}$ and $M_{\rm total,ref}=10^5\,M_{\odot}$ on a pre-computed rectangular grid with size of $200\times 200\times 1250$ for$\Phi_0\in[1.4,16]$, $g\in[0.001,3.49]$, and $r\in[0,69]$ \footnote{Note that the boundaries of the grid used in the interpolator for $\Phi_0$ and $g$ are in fact broader than the priors of $\Phi_0$ and $g$ chosen in Eq.~\ref{eq:prior-Phi0} and \ref{eq:prior-g} in Sec.~\ref{sec:prior}, so our emulator will produce accurate DFs within the prior range. The maximum value for the $r$ grid, 69, is chosen to be greater than ${\rm max}\{\hat{r}_{\rm t}(\Phi_0,g)|\Phi_0\in[1.4,16]$, $g\in[0.001,3.49]\}$, which is the maximum of the normalized truncation radius $\hat{r}{\rm_t}$ within the grid of $\Phi_0$ and $g$.} variables respectively using the \texttt{limepy} code,
    \item obtain the true values of $A$, $r{\rm_t}$, $s$, and $\phi$ at certain $r_{\rm h}$ and $M_{\rm total}$ with the scaling relations in Eq.~\ref{eq:scaling}, and
    \item substitute the interpolated and re-scaled values of $A$, $r{\rm_t}$, $s$, and $\phi$ into the analytical functions in Eq.~\ref{eq:DF}-\ref{eq:Egamma} to calculate the limepy DF $f_{\rm lp}(x,v|\boldsymbol{\theta}_{\rm lp})$.
\end{enumerate} 
Our emulator is now entirely based on analytical functions, and the gradient of the emulator with respect to any parameters can then be reliably obtained with AD in \texttt{JAX}. For functions with only two or three variables, linear interpolation is both accurate and practical with well-understood error properties compared to more complicated approaches that might rely on, e.g., neural networks.

To show the performance and accuracy of our interpolation-based emulator, we compare evaluations of the logarithmic lowered isothermal DF $\log f(x,v)$, which was the quantity used in MCMC, from both the original \texttt{limepy} code and our emulator. As seen in Fig.~\ref{fig:emulator-accuracy}, our emulator agrees well with the original function and has errors on the order of or smaller than $0.1\%$ across different parameter combinations and ranges. In particular, our emulator reproduces accurate DFs even when $g$ and $\Phi_0$ are close to the boundary values used in the interpolation, as seen in Fig.~\ref{fig:emulator-accuracy-1041}. By using JIT in JAX and using interpolation to avoid solving ODEs, we achieve a speed-up of more than 50 times for the likelihood evaluation step. Evaluating the DF from \texttt{limepy} code with a given set of $(\Phi_0,g,M_{\rm tot},r_{\rm h},\vec{r},\vec{v})$ parameters takes about 0.06 s, while the same evaluation from our emulated code takes only about 0.001 s. This speed-up enables our HMC sampling to be completed in about one or two hours instead of days on a single CPU. 

To show that our emulator is sufficiently accurate for performing Bayesian inference, we use both the \texttt{limepy} code and our emulator to perform the non-hierarchical Bayesian inference of GCs introduced in EWR22. Assuming complete and accurate knowledge for both the phase-space information of all $N$ stars in the GC and the phase-space center of the GC, we want to infer $\boldsymbol{\theta}_{\rm lp}$, the four structural parameters of the GC. The likelihood of this simpler model is given in Eq.~\ref{eq:limepy-df-likelihood}, while we choose Eq.~\ref{eq:prior-Phi0}-\ref{eq:prior-rh} as the model prior, which is the same prior for our hierarchical model. To also test the performance of HMC, we use slice sampling \citep{Slice}, a popular alternative to the Metroplis-Hasting algorithm for non-hierarchical Bayesian inference, when we run the \texttt{limepy} code for likelihood evaluation; we rely on HMC sampling when our emulator is used to calculate the lowered isothermal DF. The results of MCMC runs under these two settings for the two perfect simulations are shown in Fig.~\ref{fig:basic_Bayesian}. We see that the inference results using the emulator+HMC method agree with the baseline results from the \texttt{limepy} code+slice sampling method in the two cases shown in Fig.~\ref{fig:basic_Bayesian} (the overlapping blue and red contours), demonstrating the effectiveness and accuracy of our emulator for the \texttt{limepy} code. The \texttt{limepy} code+slice sampling method takes about an hour to sample 4000 points in a single chain for inferring the four GC structural parameters, while running HMC with our emulated code takes only about two to three minutes for sampling the same number of points. In addition, we also wish to highlight the partial degeneracy between $\Phi_0$ and $g$ based on the inference results shown in Fig.~\ref{fig:basic_Bayesian}: a decrease in $g$ can be partially compensated by an increase in $\Phi_0$ to preserve the GC morphology.

\section{Results}\label{sec:results}

We report the inference results on the simulations described in Sec.~\ref{sec:simulated-data} using our hierarchical model. Given that the likelihood is defined by the DF that was used to generate the data, we expect to obtain reasonable parameter estimates through inference made from the posterior distribution using MCMC. However, given the complexity of our models ($\sim$6010 parameters, numerical DFs, and non-trivial coordinate transformations) and the use of techniques including emulation, auto-differentiation, and HMC, it is critical to validate our model performance on simulations, in which case the true parameters are known. We are also going to impose prior distributions that are at least weakly informative, and so it is good practice to test whether the posterior can still be used to reliably infer the model parameters. Furthermore, we can explore the impacts of different observational settings on inference results through changing the hyperparameters (e.g. number of stars observed $N$, distance of cluster centers $R_{\rm c}$) described in Table~\ref{tab:sim-obs-param}, which further validates the performance of our model. 

\begin{figure*}
\begin{subfigure}{0.98\textwidth}
\centering
	\includegraphics[width=\columnwidth]{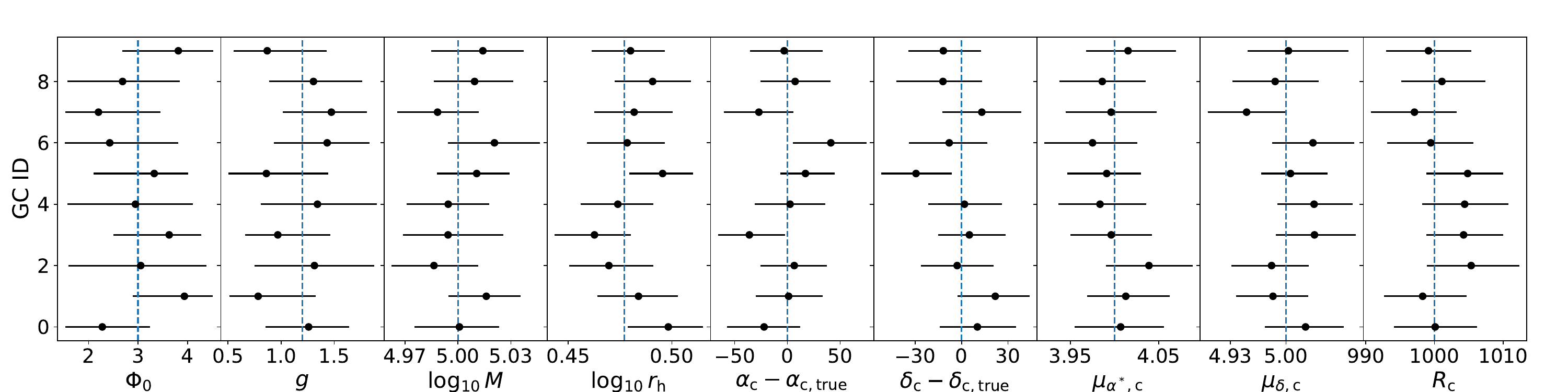}
     \label{fig:CI-240}
     \caption{True parameters $(\Phi_0,g,\log_{10}M_{\rm tot},r_{\rm h})=(3,1.2,5,3)$}
 \end{subfigure}
\begin{subfigure}{0.98\textwidth}
\centering
 \includegraphics[width=\columnwidth]{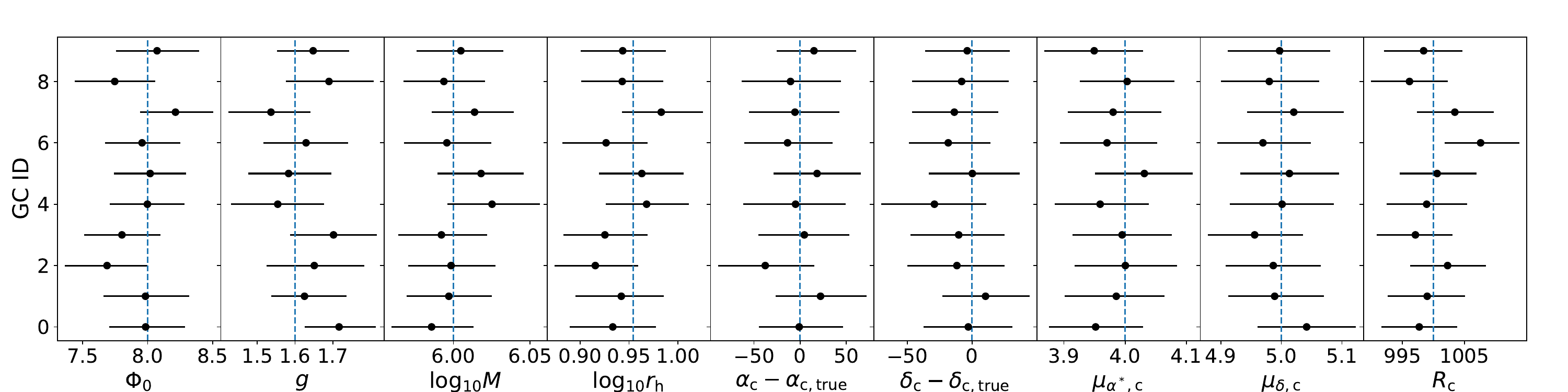}
  \label{fig:CI-270}
   \caption{True parameters $(\Phi_0,g,\log_{10}M_{\rm tot},r_{\rm h})=(8,1.6,6,9)$}
\end{subfigure}
\caption{The parameter estimates and the 95$\%$ credible intervals for two different sets of structural parameters $\lps$ (out of 32), with 10 simulations generated for each parameter set. Each panel shows the 10 credible intervals (error bars corresponding to 95$\%$ probability of the posterior), the corresponding mean (black points), and the true parameter value (vertical blue line). The vertical axis gives the ID of the simulations within the same parameter set, while the horizontal axis of each panel gives the range of each GC parameter. The angular position parameters $\alpha_{\rm c}$ and $\delta_{\rm c}$ are in the unit of arcsec, and the proper motion parameters are in mas/year. We plot the difference between the sampled values and the true values $\alpha_{\rm c}-\alpha_{\rm c, true}$ and $\delta_{\rm c}-\delta_{\rm c, true}$ for angular positions. Our hierarchical model recovers the underlying true parameters well. Note that in both the top and bottom panels, the 95\% credible intervals overlap the true value 90-100\% of the time, for all parameters. This indicates that the 95\% credible intervals are reliable.} 
\label{fig:credible-interval}
\end{figure*}

\begin{figure*}
\centering
\includegraphics[width=1.85\columnwidth]{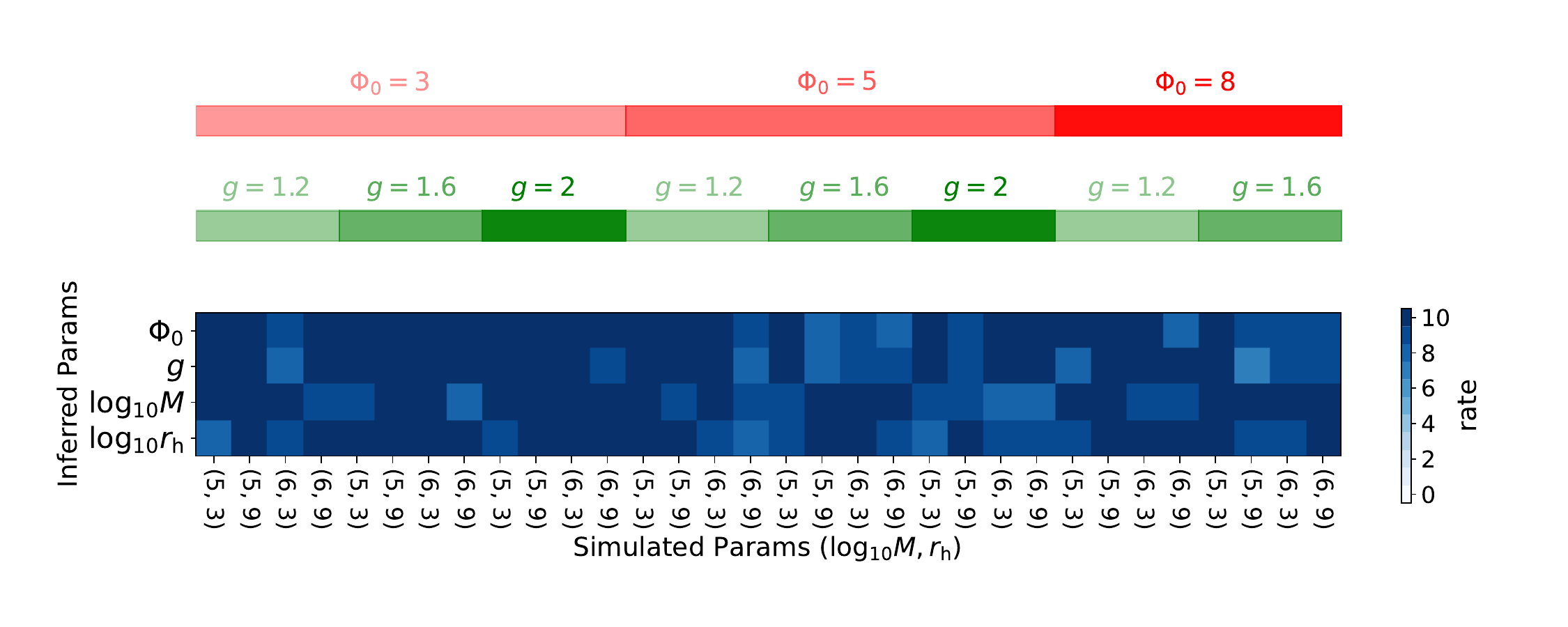}
\caption{This figure summarizes the number of times the true $\lps$ parameter value fell within the infered 95$\%$ credible interval. The vertical axis on the left indicates the parameters, and the horizontal axis along the bottom indicates the underlying true $(\log_{10}M,r_{\rm h})$ values used for each set of the 10 simulations. The red and green colour bars at the top indicate the true $(\Phi_0,g)$ values used for each set of simulations. The blue colours in the figure indicate the number of times out of 10 (i.e., the rate, shown in the vertical axis on the right) that the 95$\%$ credible intervals recover the true structural parameters. For most parameter combinations, the credible interval recovers the true parameter value 9 or 10 times out of the 10 simulations, as expected for the $95\%$ credible intervals. To help the reader understand this figure, we give the following example: \textit{The results for the simulation with true values $(\Phi_0=3,g=1.6,\log_{10}{M}=5, \log_{10}{r_h}=3)$ are the blue coloured boxes in the fifth column from the left. For this particular set of parameter values, the $\Phi_0, g,$ and $\log_{10}{r_h}$ parameters were recovered 10/10 times, and the $\log_{10}{M}$ parameter was recovered 9/10 times.}}
\label{fig:coverage-rate}
\end{figure*}

\subsection{GC Structural Parameters}\label{sec:result-GC-structure}
Here we analyze and summarize the inference results of the $32\times 10$ simulations for the 32 sets of structural parameter values, which were given in Table~\ref{tab:sim-GC-param} of Sec.~\ref{sec:simulated-data}. With the same underlying structural parameters, the 10 simulations are different realizations of the same distribution function achieved through random sampling of stars. Using one CPU per chain, it takes about one to two hours for HMC to sample 4000 points in each chain under our hierarchical model, which is significantly more time-consuming than sampling the non-hierarchical model described in Sec.~\ref{sec:emulation}. Further speed-up can be achieved by using multiple CPUs per chain or GPUs for DF evaluations and sampling. By default, we run four chains for each HMC run and discard the first $50\%$ of the samples for burn-in and adaptive tuning.

We notice that one or two out of the four chains can get stuck in some local minimum (four chains do not agree), which occur roughly one fourth of the time out of all HMC runs. One particular noticeable local minimum exists around $\Phi\approx 14$, where the sampling often gets stuck. Rerunning HMC one or two times can usually fix the problem and produce converging inference results for all four chains. However, there are 10 simulations (out of 320) where one chain is persistently trapped in some local minimum (disagreeing with the other three chains) even after multiple reruns of HMC. For these 10 simulations, we simply discard the chain stuck in the local minimum and keep the other three chains (which are consistent with each other) as the inference results.

In addition, the NUTS algorithm in \texttt{PyMC} fails to produce any results for 3 simulations (out of 320). In all three cases, the adaptive tuning of the mass matrix fails and the diagonal of the mass matrix contains zero value, which halts the sampling process. Since this problem occurs relatively infrequently, we simply replace these three simulations with new simulations generated under different random seeds. We will address the local minimum problem and the adaptive tuning failure of NUTS in future works. 

We also notice that when the simulations were generated with $\Phi_0=2$ (not as part of the $32\times 10$ simulations analyzed in this section), which is close to the lower boundary 1.5 for the prior of $\Phi_0$, the sampled posterior of $\Phi_0$ can become overly non-Gaussian with features such as small peaks or oscillations in the shape of the 1D probability density function, despite the accuracy of our emulated DF at $\Phi_0=1.5$ as shown in Fig.~\ref{fig:emulator-accuracy-1041}. To address this sampling issue for $\Phi_0$ values near the boundaries, we might need to replace the uniform priors in Eq.~\ref{eq:prior-Phi0} and \ref{eq:prior-g} with smooth distributions that do not possess sharp boundary cutoffs.

In general, our HBI gives reasonable parameter estimates for all 320 simulations outlined in Table~\ref{tab:sim-GC-param}. We show the 95$\%$ credible intervals of the GC parameters for two different sets of structural $\lps$, as examples, in Fig.~\ref{fig:credible-interval}. We summarize the inference results of all 320 simulations in Fig.~\ref{fig:coverage-rate} by plotting the parameter recovery rate on a blue scale. This is an estimation of the frequentist coverage probability of our Bayesian credible intervals; it represents the number of times out of 10 that the 95$\%$ credible interval overlaps the true $\lps$ values for each parameter set. The $95\%$ credible interval recovers the true parameter value 9 or 10 times out of the 10 simulations in most cases.

Overall, we take these results to mean that our emulator for \texttt{limepy} is working well, and that except for perhaps some extreme cases, the credible regions are reliable. In general, Fig.~\ref{fig:credible-interval} and \ref{fig:coverage-rate} demonstrate that our hierarchical model can give robust and reliable estimates of GC properties across different values for GC structural parameters.

\subsection{Hyperparameters}\label{sec:hyper-param}

Here we explore the impact of different observational conditions on the inference results to further showcase the performance of and validate our hierarchical model. The different observational conditions are reflected by the hyperparameters listed in Table~\ref{tab:sim-obs-param}. Note that we use the same structural parameters for $\lps$ and the same random seed to generate all the simulations analyzed in these section.
\begin{figure*}
	\includegraphics[width=1.7\columnwidth]{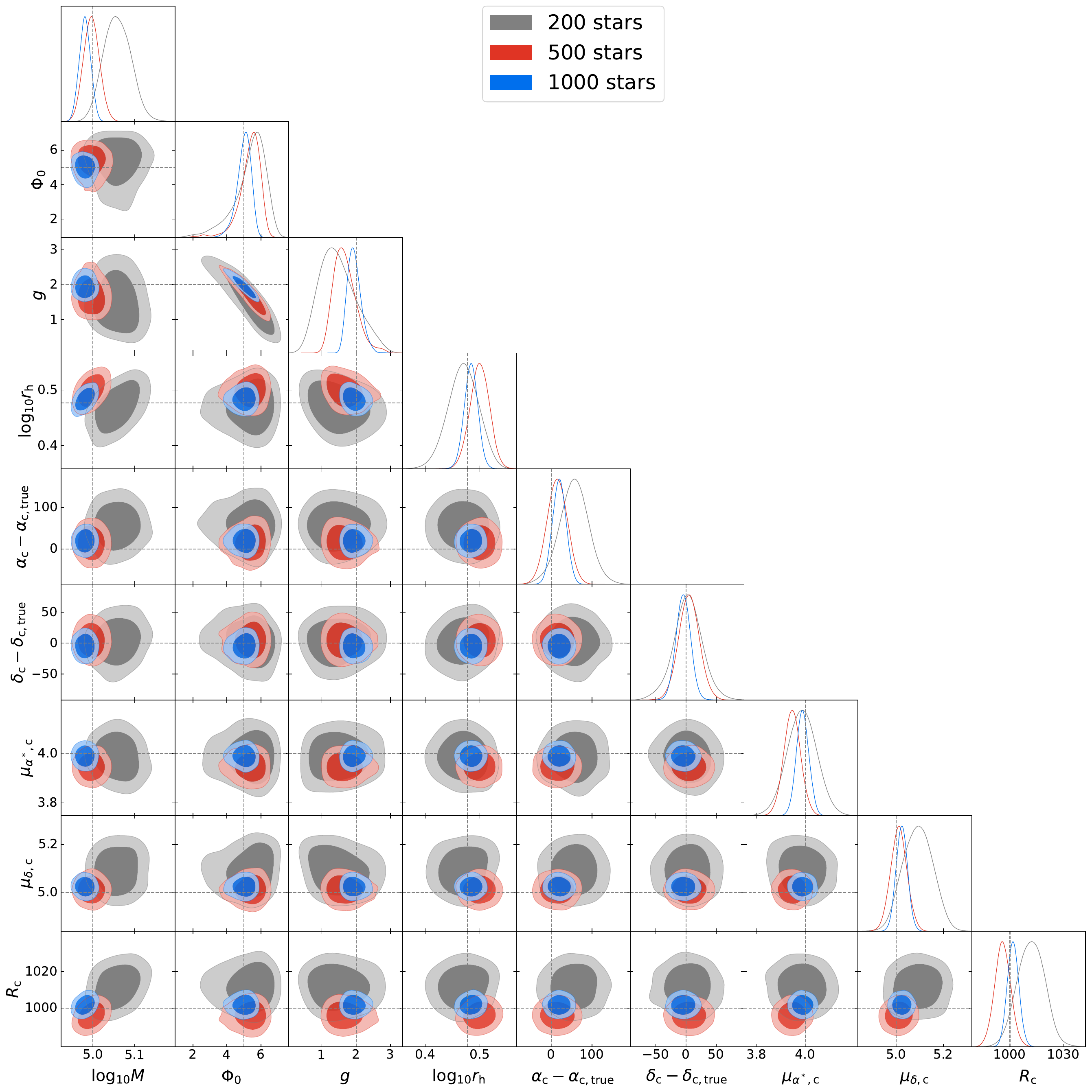}
    \caption{The impact of different $N$, the number of observable stars in GC, on inference results. For this and all the following figures, we use the dark and the ligher color to indicate the $75\%$ and $95\%$ credible regions of the parameter posteriors respectively. The difference between the sampled angular position parameters and their true values, $\alpha_{\rm c}-\alpha_{\rm c,true}$ and $\delta_{\rm c}-\delta_{\rm c,true}$, are shown in the unit of arcsec. Increasing the number of stars clearly improve the inference results for all GC parameters.}
    \label{fig:change_number_of_stars}
\end{figure*}
 
Increasing the number of observable stars in a GC significantly improves the constraints for all GC parameters (Figure~\ref{fig:change_number_of_stars}). At the same time, we note that meaningful and relatively accurate estimates of the GC properties can still be obtained with only 200 uniformly selected stars due to the use of both spatial and kinematic information in our model. 

\begin{figure*}
\begin{subfigure}{.47\textwidth}
\centering
	\includegraphics[width=\columnwidth]{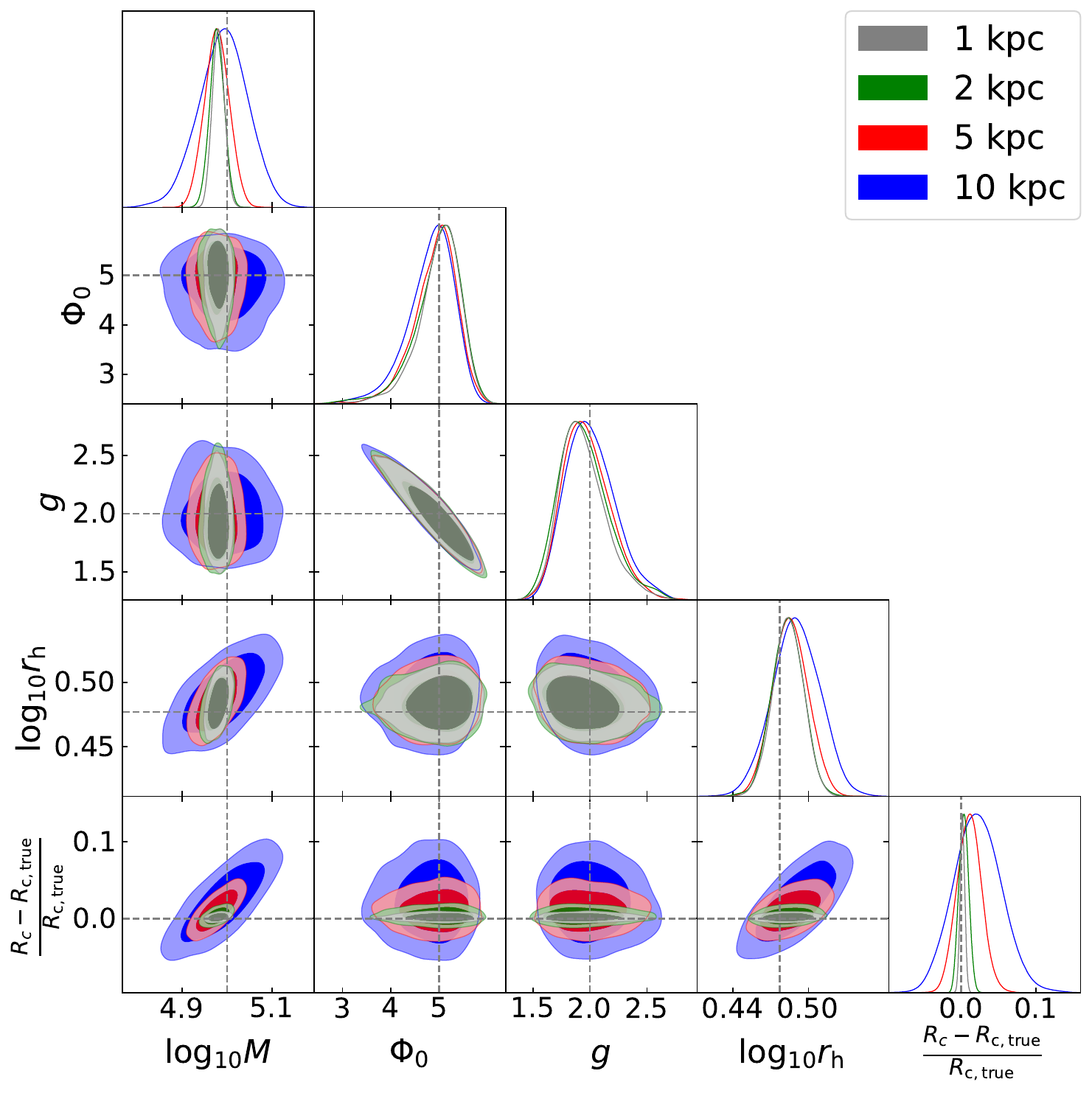}
  \caption{Estimates for the structural parameters and the radial distance}
  \label{fig:change_distance_structures}
 \end{subfigure}
\hfill
\begin{subfigure}{.47\textwidth}
\centering
 \includegraphics[width=\columnwidth]{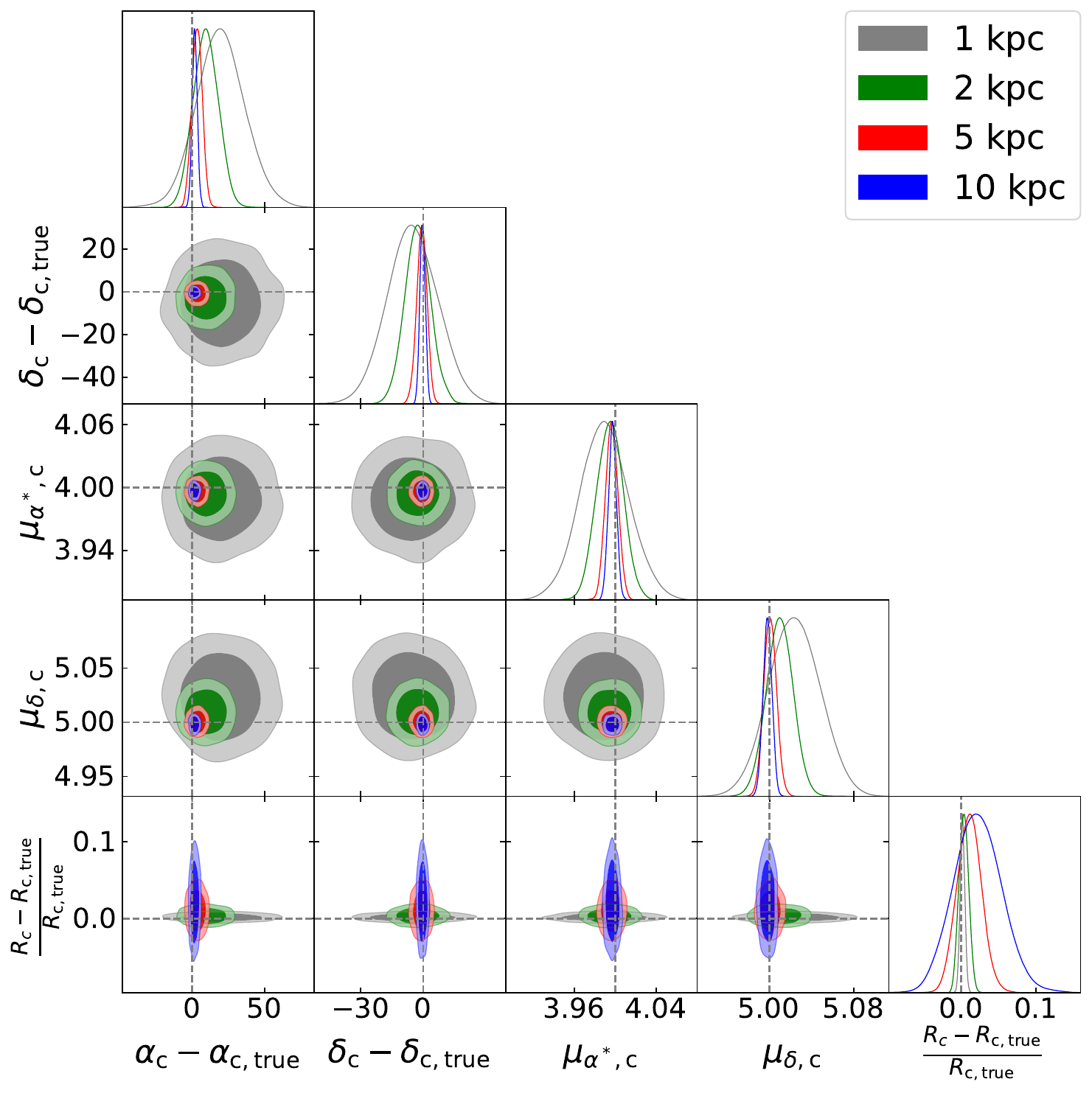}
   \caption{Estimates for the GC center parameters}
   \label{fig:change_distance_positions}
\end{subfigure}
\caption{The impact of different $R_{\rm c, true}$, the true radial distance to GC, on the inference results. We use the difference between the observed and the true distances normalized by the true distance, $(R_{\rm c}-R_{\rm c, true})/R_{\rm c, true}$, to indicate the uncertainty on the radial distance estimate. In the left subplot, we see that the uncertainty of radial distance has a noticeable impact on the uncertainty of the mass estimate and, to a less extent, the half-mass radius estimate for the simulated GC. A lower radial distance measurement correlates with a lower mass and a lower radius estimates for GCs with a radial distance at 5 and 10 kpc. The uncertainty of the angular positions and proper motions significantly increases with a decreasing distance, but there is no obvious correlation between the radial distance and the angular coordinates, as seen in the right subplot.}
\label{fig:change-distance}
\end{figure*}

In Fig.~\ref{fig:change-distance}, we show the inference results for simulations placed at different radial distances $R_{\rm c}$. We see that the posterior distributions on $\Phi_0$, $g$, and $r_{\rm h}$ are relatively insensitive to the change of $R_{\rm c}$, while increasing distance noticeably enlarges the uncertainty of the mass estimate (the first column of Fig.~\ref{fig:change_distance_structures}). This behavior is expected; the morphology and shape of GCs are relatively robust to the measurement error on stars' positions and velocities in the Cartesian coordinates when we have hundreds of stars, while $M\propto v^2$ is sensitive to the error on $v$, which depends on the observer's distance to GC. With the same uncertainties for the angular proper motions of individual stars, a larger radial distance $R_{\rm c, true}$ leads to a larger error on $v^2$ and therefore a larger error on $M$. In addition, there is a noticeable degeneracy between the GC radial distance and the GC mass (seen in the lower left corner of Fig.~\ref{fig:change_distance_structures}): a lower radial distance estimate leads to a lower mass estimate, which suggests that any bias in estimating the radial distance to the cluster center can lead to a bias in estimating the GC mass. The measurement of the half-mass radius $r_{\rm h}$ is also somewhat degraded, especially when $R_{\rm c, true}$ increases from $5$kpc to $10$kpc. This suggests the importance of including the radial distance parameter $R_{\rm c}$ into the model, and our hierarchical model provides the clear benefit of robustly including the uncertainty of the parallax of each individual star, thereby inferring the GC mass, GC radius, and the radial distance to the GC center at the same time and properly accounting for the degeneracy among these parameters.

In Fig.~\ref{fig:change_distance_positions}, the uncertainty of the angular GC center parameter decreases as the distance increases, which is expected since a GC with a fixed physical size occupies a smaller area of the sky as its distance increases. However, there is no correlation between the radial distance estimate and the estimates of the angular coordinates and proper motions, which is evident from the last row of Fig.~\ref{fig:change_distance_positions}. As shown in the 1D posterior plot (the lower right corner of both subplots in Fig.~\ref{fig:change-distance}) for the $(R_{\rm c}-R_{\rm c, true})/R_{\rm c, true}$ parameter, which indicates the relative measurement error of the radial distance to the GC center, our HBI procedure can reliably recover the distance of the cluster center regardless of distance, but that the relative precision of the cluster distance estimate decreases as the true distance increases. This behaviour reflects the fact that the same parallax measurement errors become less informative as we increase the radial distance (decrease the parallax value). 

\begin{figure*}
\begin{subfigure}{.47\textwidth}
\centering
	\includegraphics[width=0.9\columnwidth]{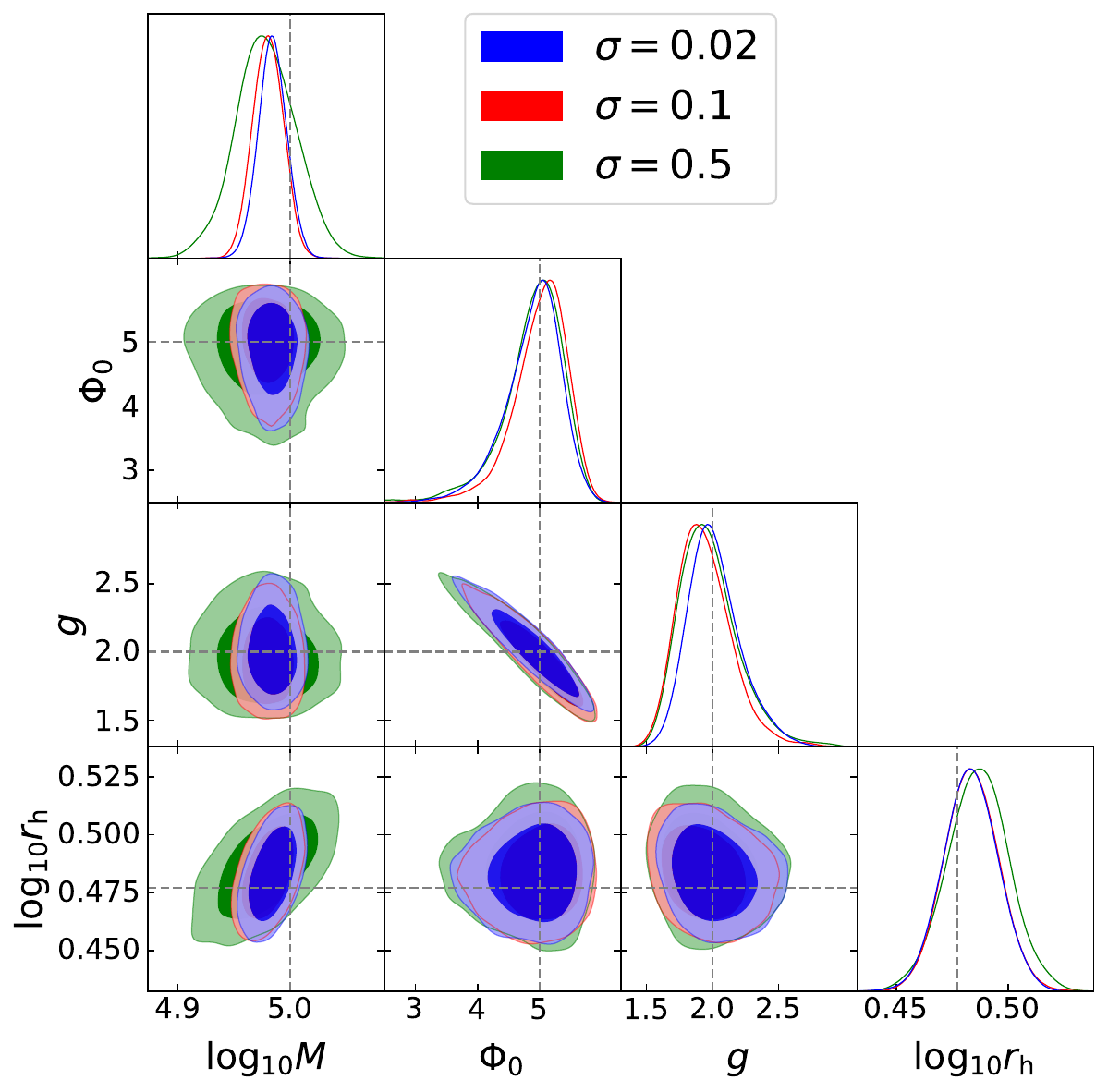}
     \caption{Estimates for the structural parameters}
     \label{fig:change_measure_structures}
 \end{subfigure}
\hfill
\begin{subfigure}{.47\textwidth}
\centering
 \includegraphics[width=0.9\columnwidth]{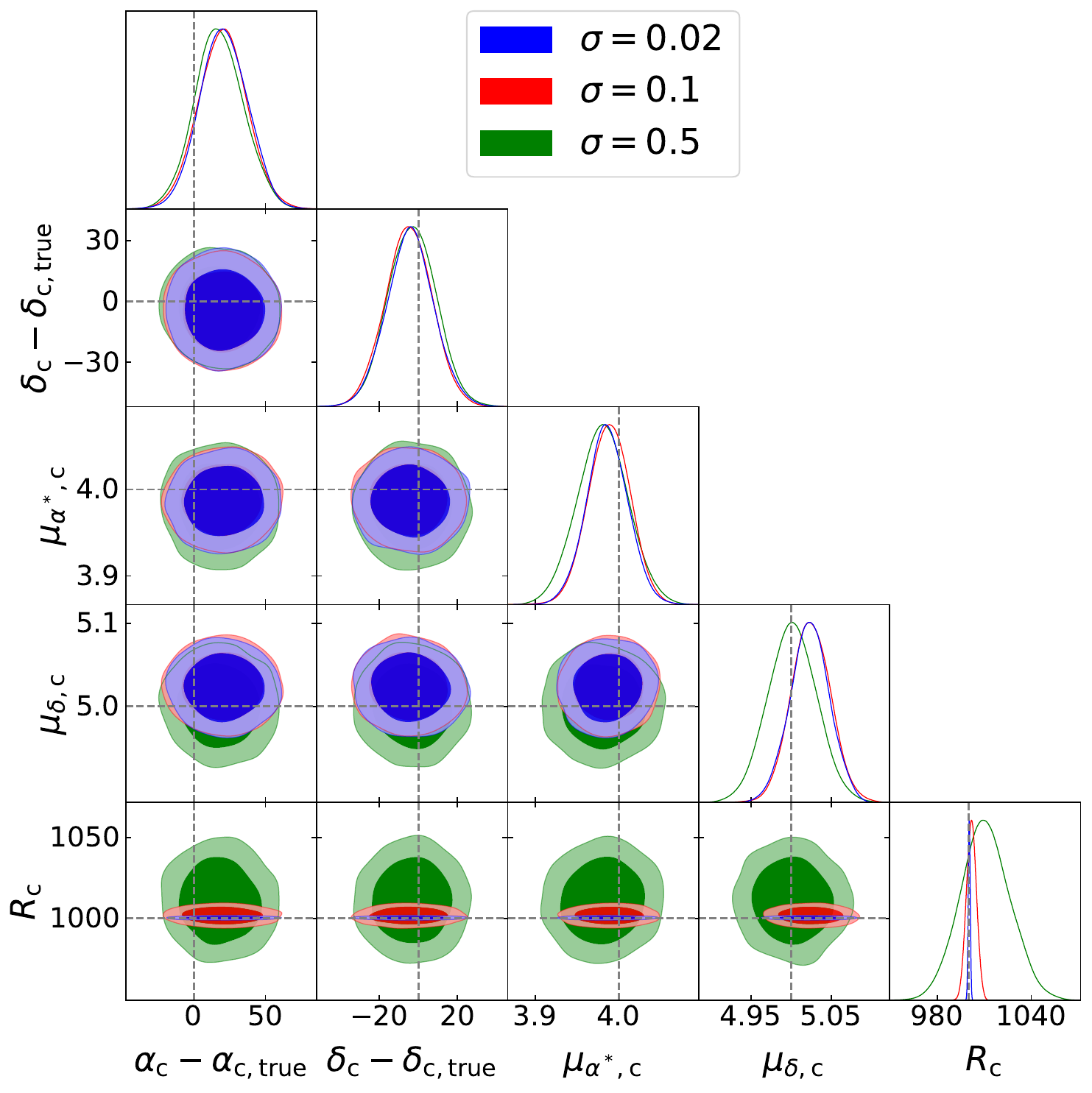}
   \caption{Estimates for the GC center parameters}
   \label{fig:change_measure_positions}
\end{subfigure}
\caption{The impact of different $\sigma$, measurement errors of angular positions, parallax, and proper motions, on inference results. The uncertainty of the mass estimate noticeably increases when the measurement errors increase from $0.1$ to $0.5$ (in units of mas or mas/year). The uncertainty of the radial distance estimate increases proportionally with respect to the measurement errors.}
\label{fig:change-measure}
\end{figure*}

We next show the impact of different $\sigma$, measurement errors of angular positions, parallax, and proper motions, in Fig.~\ref{fig:change-measure}. For the structural parameters presented in Fig.~\ref{fig:change_measure_structures}, an increase of measurement errors from $0.02$ to $0.1$ (in units of mas or mas/year) make little difference to the parameter constraints, while the uncertainty of the mass estimate noticeably increases when the measurement errors increase from $0.1$ to $0.5$. The constraints of GC angular positions and proper motions are relatively insensitive to the measurement errors (Fig.~\ref{fig:change_measure_positions}), while the uncertainty of the radial distance estimate increases proportionally with respect to the measurement errors, mostly due to the dependence of radial distance estimates on the measurement errors of parallax. 

\begin{figure*}
\begin{center}
	\includegraphics[width=1.8\columnwidth]{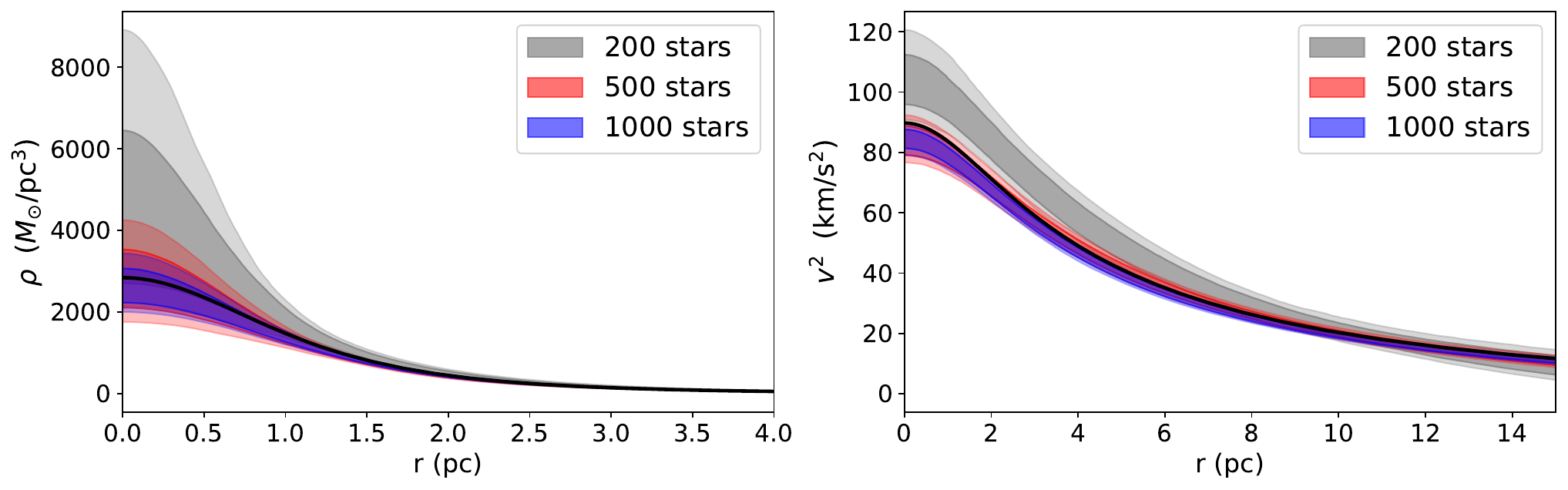}
\end{center}
    \caption{The constraints of the density $\rho$ and velocity dispersion $v^2$ profiles from HMC runs on simulations with different numbers of stars $N$. The solid black curves show the actual density profile evaluated at the true structural parameters that are used to generate these simulations. The constraints on both profiles noticeably improve with the inclusion of more number of stars.}
    \label{fig:density-v2-number}
\end{figure*}

\begin{figure*}
\begin{center}
	\includegraphics[width=1.8\columnwidth]{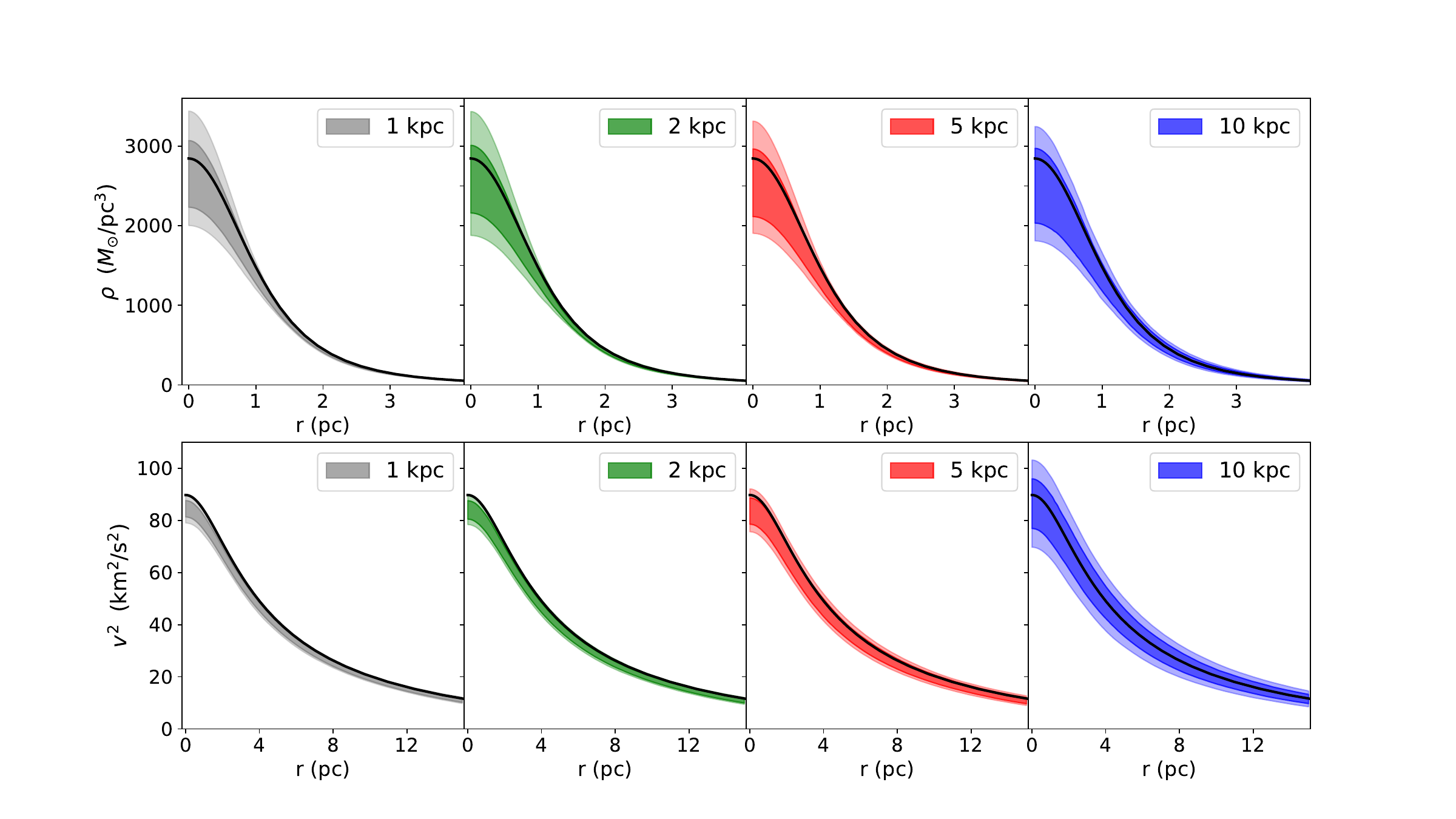}
\end{center}
    \caption{The constraints of the density $\rho$ and velocity dispersion $v^2$ profiles from HMC runs on simulations with different radial distance $R_{\rm c}$. The solid black curves show the actual velocity dispersion profile evaluated at the true structural parameters that are used to generate these simulations. We set the y axis with the same range for all the panels in each row to allow easy comparison. The constraints for $\rho(r)$ are similar for different $R_{\rm c}$, while the uncertainties for $v^2$ increase as $R_{\rm c}$ increases.}
    \label{fig:density-v2-distance}
\end{figure*}

\begin{figure*}
\begin{center}
	\includegraphics[width=1.4\columnwidth]{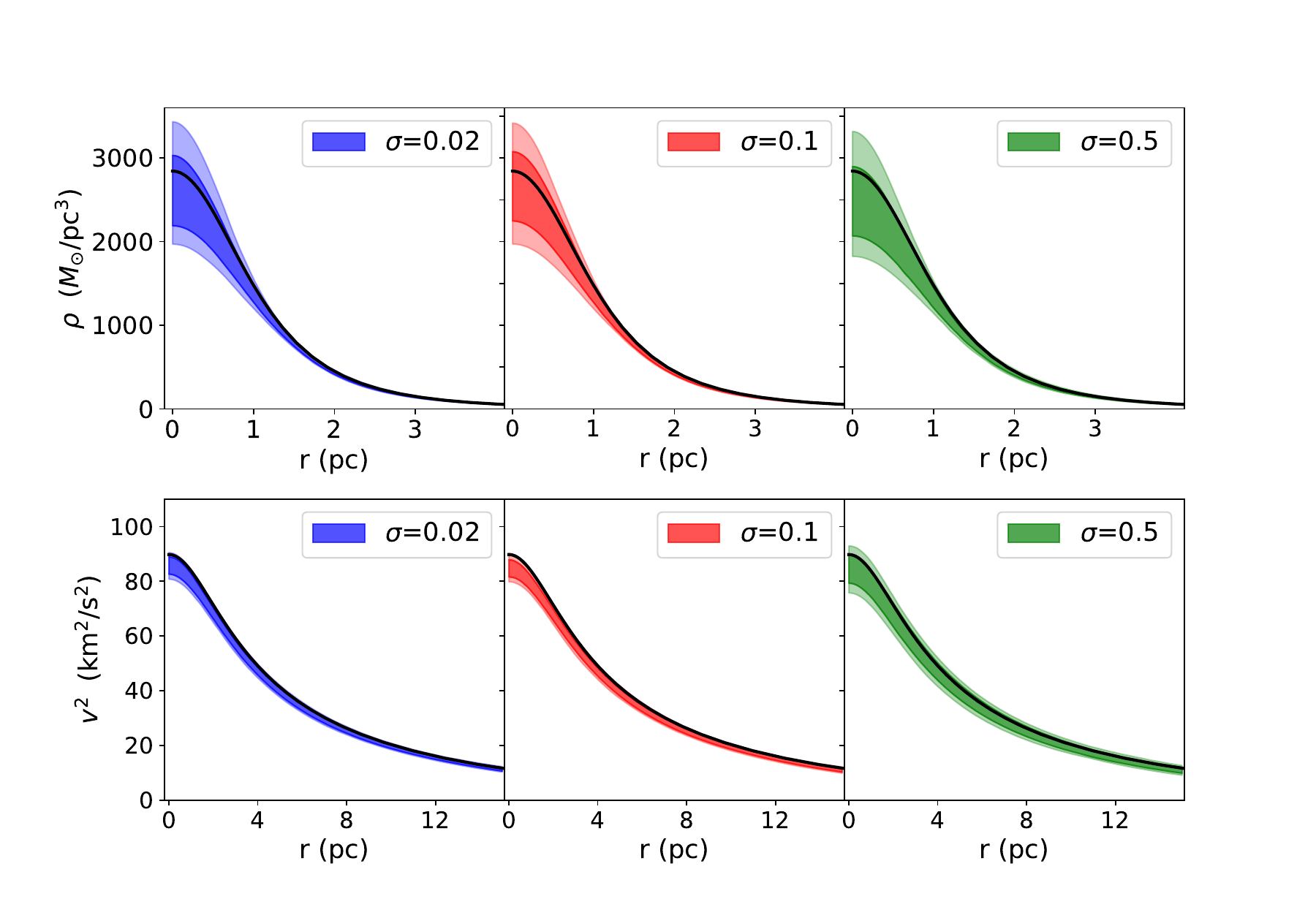}
\end{center}
    \caption{The constraints of the density $\rho$ and velocity dispersion $v^2$ profiles from HMC runs on simulations with different measurement uncertainties $\sigma$.}
    \label{fig:density-v2-error}
\end{figure*}

By utilizing the posterior distribution estimation of a single GC, we can not only estimate parameters but also functions such as the density profile and the cumulative mass profile. The procedure we use to estimate the density profile is the same as the one outlined in \cite{19EadieGCMW} for estimating the Milky Way's cumulative mass profile. We apply this procedure to each set of GC structural parameters sampled through HMC to calculate the density profile $\rho(r)$ using the \texttt{limepy} code. As we have thousands of samples in our Markov chain, we generate thousands of density profile estimates. These estimates offer both a visual and quantitative representation, enabling us to compute Bayesian credible regions and to compare the estimates directly to the GC's true density profile. Another quantity of interest that we can estimate using the posterior distribution for a single GC is its velocity dispersion.

We plot the density and velocity dispersion profile constraints of the simulated GCs for different number of stars $N$ in Fig.~\ref{fig:density-v2-number}, different distances $R_{\rm c}$ in Fig.~\ref{fig:density-v2-distance}, and different measurement errors $\sigma$ in Fig.~\ref{fig:density-v2-error}. For these selected clusters, we recover both the density and velocity dispersion profiles very well through our HMC inference. The accuracy of density and velocity dispersion profiles reflect the accuracy of our constraints on $\boldsymbol{\theta}_{\rm lp}$: in Fig.~\ref{fig:density-v2-number}, we see that the constraints of both density and velocity dispersion profiles are significantly improved when the number of stars increases, which increases the amount of available information about the GC and thereby improves the constraints on $\boldsymbol{\theta}_{\rm lp}$. In Fig.~\ref{fig:density-v2-distance} and \ref{fig:density-v2-error}, the constraints of $\rho(r)$ are comparable for different  $R_{\rm c}$ and $\sigma$, while the uncertainties for the velocity dispersion profiles $v^2$ noticeably increase as $R_{\rm c}$ and $\sigma$ increase. Since $M\propto v^2$, the amplitude of the velocity dispersion profiles $v^2$ indicates the value of $M_{\rm total}$. The uncertainties in $v^2(r)$ therefore reflect the degradation of constraints on $M_{\rm total}$, illustrated in Fig.~\ref{fig:change_distance_structures} and \ref{fig:change_measure_structures}.

In general, our hierarchical model performs well for simulations with different hyperparameters reflecting different observational conditions. The samples approximating the posterior distribution both (1) recover the underlying true parameters, and (2) give reasonable uncertainty estimates that are consistent with the assumed observation conditions.  Together, these two achievements demonstrate the power of HBI to incorporate measurement errors and missing data into parameter estimation.

\section{Conclusions}\label{sec:conclusions}

Using the lowered isothermal DF proposed by GZ15, we build a hierarchical Bayesian model for GC inference that can directly take data in the heliocentric equatorial coordinates (the reference frame in which we obtain measurements) and can incorporate incomplete data and measurement errors. We extend the basic Bayesian modelling of GCs in EWR22 into the hierarchical version and further include the inference for the positions and velocities of the GC center into the hierarchical model. To render the inference process computationally feasible, we build an interpolation-based emulator for the single-mass, lowered isothermal DF, written in \texttt{JAX}, which allows auto-differentiation, and we run Hamiltonian Monte-Carlo to estimate the posterior. We test our hierarchical model on hundreds of GC simulations with different parameters. We show that our hierarchical model can give accurate and reliable estimates of GC properties across different GC morphologies and distances, and our approach gives robust uncertainty estimates that incorporate measurement errors and are consistent with the assumed observation conditions.

There are still several technical problems related to running HMC that need to be addressed in order to allow efficient and accurate hierarchical Bayesian inference on the entire parameter range. In addition to addressing the local minimum problem and updating the uniform priors with smoother priors to avoid sharp boundary cutoffs, which we have already discussed in Sec.~\ref{sec:result-GC-structure}, we also face difficulty with using HMC to sample posteriors for simulations with $g\lesssim0.6$: the chains fail to converge due to issues with adaptive tuning and produce unstable inference results. Even though most of the Galactic GCs have been fitted with $g\geq 0.8$ where our code does produce reliable results, there is still a sizable fraction of GCs fitted with $g\lesssim0.6$ \citep{19deBoerGCNumberDensityGaiaDR2,23Dickson_IMF}. It is therefore important to continue our investigation into the parameter specification of our model and the auto-tuning procedures adopted in NUTS so that we can perform HBI in the entire prior range of $(\Phi_0,g)$.

Another technical area for further efforts is to improve the speed of the HMC sampling. Currently, it takes one to a few hours to sample hierarchical models with 6010 parameters for 1000 stars, using one CPU per chain during HMC. This speed is sufficient for including stars with kinematic information, which range from $O(10)-O(10^3)$ number of stars for GCs in the Gaia data \citep{19Vasiliev_GaiaDR2_pm}. However, in a real GC there can be many more stars with known angular positions but without any kinematic information. To include all these stars, one would need to infer very large models with $O(10^4)$ numbers of stars, which renders the speed-up of HMC necessary. Such speed-up can be achieved by just-in-time compilation, GPU-acceleration, and further parallelization of the HMC sampler.

For the likelihood of our model in Eq.~\ref{eq:limepy-df-likelihood}, the independence and uniformity assumption does not take into account of the selection bias occurred in real surveys such as Gaia and HST due to source-crowding, magnitude limits, the contamination of field stars, and other observational effects. EWR22 has emphasized the importance of selection function modelling for real survey data on GCs in order to avoid any severe inference bias. Such selection function modelling is naturally addressed in a hierarchical framework, and we plan to extend our model for observed data in future work. This is especially relevant for cases where biased sampling also requires combining observations from different datasets. However, our HBI model can already be used to fit real observational data when the sampling of stars across the clusters are uniform, and we aim to apply our HBI model to GC data compiled in \citet{19deBoerGCNumberDensityGaiaDR2} in future work.

With the ability to properly address selection functions and combine different datasets, the natural next step to extend and test our hierarchical model will be to include radial velocity measurements, which are available for a sizable number of stars in a large fraction of Milky Way GCs from ground-based spectroscopy \citep{22Apogee,22LAMOST}. It is also important to compare the method presented here to traditional methods in the literature that (1) use the projected distances of stars to estimate density and mass profiles with radial binning, and (2) combine data sets from different telescopes to use stars at all radii. 

We have so far only been able to incorporate the single mass, isotropic, lowered isothermal DF into our hierarchical model, which limits the applicability and flexibility of the model and can lead to significant biases on inferred GC properties such as mass and radius due to ignoring dynamical effects such as mass segregation \citep{15SollimaBias,19Brunet_M4Nbody}. In order to explore interesting GC features such as dark remnants, binaries, and mass segregation and to achieve more realistic GC modelling, future work should focus on building emulators for the multi-mass, anisotropic, lowered isothermal GC model, therefore allowing the hierarchical inference to utilize the full power of the \texttt{limepy} code. However, building emulators of the anisotropic, multi-mass model is more difficult due to a substantial increase in the complexity of the model and the number of parameters, which makes the linear interpolation process less efficient, if not impossible. One can perhaps build upon the simple interpolation approach laid out in Sec.~\ref{sec:emulation} for the anisotropic, multi-mass model or use emulators based on machine learning. Alternatively, one could rewrite the full \texttt{limepy} code to make it auto-differentiable so that we can obtain reliable gradients for HBI. 

Astronomers are interested not only in the inherent characteristics of globular clusters (GCs), but also in comparing and selecting GC models. The latter is crucial for comprehending the internal dynamics of GCs and the broader narrative of their evolution as they move through the Galactic potential. Being able to associate an observed GC with a particular dynamical model is a critical step in unraveling the cluster's present properties and its evolutionary past. Understanding the distribution function of stars within a cluster enables more intricate features to be explored in greater detail, like its dark remnant population, binary population, degree of mass segregation, and tidal history. By utilizing a model that integrates all of these components and also enhancing the statistical framework to account for observational sampling bias, astronomers can gain a better understanding of the dynamical state of GCs. Knowing a cluster’s dynamical state also places constraints on the cluster’s properties at birth and how it has evolved over time. By demonstrating the robustness and practicality of HBI, our work lays out the foundation for more realistic and complex GC modelling that has the potential to maximize the cluster’s utility as a tool to study the universe around it.

\section*{Acknowledgements}

JSS would like to thank Rebecca Bleich for continuing to tolerate his colour preferences. JSS acknowledges funding from the Dunlap Institute and from NSERC through Discovery Grant RGPIN-2023-04849. GE acknowledges funding from NSERC through Discovery Grant RGPIN-2020-04554. To perform inference on our hierarchical model, we use \texttt{limepy} (\url{https://github.com/mgieles/limepy}), \texttt{JAX} (\url{https://jax.readthedocs.io/en/latest/}), and \texttt{PyMC} (\url{https://github.com/pymc-devs/pymc}). We use \texttt{getdist} (\url{https://github.com/cmbant/getdist}) to make corner plots that demonstrate Bayesian parameter constraints. To plot the directed acyclic graph (DAG), we use \texttt{daft} (\url{ https://docs.daft-pgm.org/en/latest/}), and we thank Jeff Shen for helping with the DAG plot.

\section*{Data Availability}
The data underlying this article will be shared on reasonable request to the corresponding author.

\bibliographystyle{mnras}
\bibliography{GC} 

\appendix

\section{Coordinate Transform}\label{sec:coordinate}
In the Heliocentric reference system, the transformations from equatorial coordinates $(\boldsymbol{q},\boldsymbol{p})$ to Cartesian coordinates $(\mathbf{x},\mathbf{v})=T(\boldsymbol{q},\boldsymbol{p})$ are given by 
\begin{align}
x& =R \cos \alpha \cos \delta \label{eq:x}\\
y & =R \sin \alpha \cos \delta \\
z & =R \sin \delta\\
v_{x} & =v_{\rm R} \cos \alpha \cos \delta-R\mu_{\alpha*} \sin \alpha-R\mu_\delta \cos \alpha \sin \delta \\
v_{y} & =v_{\rm R} \sin \alpha \cos \delta+R\mu_{\alpha*} \cos \alpha-R\mu_\delta \sin \alpha \sin \delta \\
v_{z} & =v_{\rm R} \sin \delta+R\mu_\delta \cos \delta \label{eq:vz}
\end{align}

The inverse transformation $(\boldsymbol{q},\boldsymbol{p})=T^{-1}(\mathbf{x},\mathbf{v})$ (from Cartesian to equatorial) is simply given by
\begin{align}
 R &= \sqrt{x^2+y^2+z^2} \label{eq:R}\\
 \alpha &={\rm arctan2}(y/x)\\
\delta &=\frac{\pi}{2}-{\rm arctan2}(\sqrt{x^2+y^2}/z)\\
\mu_{\alpha*}&=\frac{v_yx-v_xy}{x^2+y^2}\cos \delta\\
\mu_{\delta}&=\frac{1}{R^2}\left(v_z\sqrt{x^2+y^2}-\frac{xv_x+yv_y}{\sqrt{x^2+y^2}}z\right)\\
v_{\rm R}&=(xv_x+y_vy+zv_z)/R\label{eq:vR}
\end{align}
, where ${\rm arctan}2$ is the signed inverse of the tangent function. The Jacobian of the transformation from equatorial coordinates to Cartesian coordinates (Eq.~\ref{eq:x}-\ref{eq:vz}) is
\begin{equation}
    \left\lvert\frac{\partial T\left(\boldsymbol{q},\boldsymbol{p}\right)}{\partial(\boldsymbol{p},\boldsymbol{q})}\right\rvert=R^4\cos\delta\label{eq:Jacobian-p-q}
\end{equation}
Further transform the Heliocentric equatorial coordinates $(\boldsymbol{p},\boldsymbol{q})$ to the sampling coordinates $(\boldsymbol{s},\boldsymbol{t})$ (the relationship given in Eq.~\ref{eq:sampling-parameter}), we obtain the following Jacobian term:
\begin{equation}
   \left\lvert\frac{\partial T\left(\boldsymbol{q}(\boldsymbol{s}),\boldsymbol{p}(\boldsymbol{t})\right)}{\partial(\boldsymbol{s},\boldsymbol{t})}\right\rvert=(\frac{A}{\pi_c}+\Delta R)^4\cos\delta,\label{eq:Jacobian-s-t}
\end{equation}

\bsp	
\label{lastpage}
\end{document}